\documentclass[12pt]{article}
\usepackage{jheppub}
\usepackage{epstopdf}

\usepackage{mathrsfs}
\usepackage{psfrag}
\usepackage{color}
\usepackage{subcaption}
\usepackage{hyperref}

\usepackage{slashed}
\usepackage{feynmp-auto}
\usepackage{simplewick}
\usepackage{cancel}
\usepackage{multirow}
\usepackage[table]{xcolor}
\usepackage{amsmath,bbm,array,amsfonts,graphicx,wrapfig,arydshln,lscape,float,multirow,longtable,rotating,makecell}
\usepackage{url}

\newcommand{\smallminus}{{\rm\rule[2.4pt]{6pt}{0.65pt}}}
\newcommand{\smallplus}{\hspace{0.5pt}\text{{\small+}}\hspace{-0.5pt}}
\newcommand{\mi}{\smallminus}

\newcommand{\pl}{\smallplus}

\let\ifpdf\relax
\usepackage[T1]{fontenc}
\usepackage[utf8]{inputenc}
\usepackage{ifpdf,tocloft,rotating}
\usepackage{hyperref}
\let\normalcolor\relax
\hypersetup{pdftitle={},pdfcreator={},linkcolor=[rgb]{0.15,0.35,0.75},colorlinks=true,citecolor=[rgb]{0.675,0,0.2},urlcolor=[rgb]{0.15,0.35,0.65}}
\let\olditemize\itemize\renewcommand{\itemize}{\vspace{-2pt}\olditemize\setlength{\itemsep}{1pt}\setlength{\parskip}{0pt}\setlength{\parsep}{-0pt}}
\let\oldenumerate\enumerate\renewcommand{\enumerate}{\vspace{-4pt}\oldenumerate\setlength{\itemsep}{1pt}\setlength{\parskip}{0pt}\setlength{\parsep}{0pt}}

\renewcommand{\bar}{\overline}
\renewcommand{\hat}{\widehat}
\renewcommand{\tilde}{\widetilde}
\newcommand{\ab}[1]{\langle #1\rangle}
\newcommand{\sqb}[1]{[#1]}
\newcommand{\sab}[1]{s_{#1}}

\newcommand{\dlog}{d \log}

\newcommand{\tw}[1]{\widetilde{#1}}
\newcommand{\lra}{\leftrightarrow}

\definecolor{mhvBlue}{rgb}{0.3,0.2,0.75}
\definecolor{fRed}{rgb}{0.48,0.02824,0.18824}
\definecolor{cut2}{rgb}{0.18824,0.18824,0.48}
\definecolor{cut1}{rgb}{0.48,0.02824,0.18824}

\newcommand{\I}{\mathcal{I}}
\newcommand{\N}{\mathcal{N}}
\newcommand{\M}{\mathcal{M}}

\newcommand{\A}{\mathcal{A}}

\newcommand{\C}{\mathcal{C}}

\newcommand{\lam}[1]{\lambda_{#1}}
\newcommand{\lamt}[1]{\widetilde{\lambda}_{#1}}

\def\d{{\rm d}}

\usepackage{mathtools}

\newsavebox{\largestimage}

\title{\Large Gravity loop integrands from the ultraviolet}
\author[1,2]{Alex Edison,}
\emailAdd{alexander.edison@physics.uu.se}

\author[3]{Enrico Herrmann,}
\emailAdd{eh10@stanford.edu}

\author[1]{Julio~Parra-Martinez,}
\emailAdd{jparra@physics.ucla.edu}

\author[4]{Jaroslav Trnka}
\emailAdd{trnka@ucdavis.edu}

\affiliation[1]{Mani L. Bhaumik Institute for Theoretical Physics,\\
UCLA Department of Physics and Astronomy, Los Angeles, CA 90095, USA}

\affiliation[2]{Department of Physics and Astronomy, Uppsala University, 75108 Uppsala, Sweden}

\affiliation[3]{SLAC National Accelerator Laboratory, Stanford University, Stanford, CA 94039, USA}

\affiliation[4]{Center for Quantum Mathematics and Physics (QMAP),\\
Department of Physics, University of California, Davis, CA 95616, USA}

\abstract{We demonstrate that loop integrands of
  (super-)gravity scattering amplitudes possess surprising properties in the
  ultraviolet (UV) region. In particular, we study the scaling of multi-particle unitarity cuts 
  for asymptotically large momenta and expose an improved
  UV behavior of four-dimensional cuts through seven loops as compared to
  standard expectations. For $\N=8$ supergravity, we show that
  the improved large momentum scaling combined with the behavior of the integrand under BCFW
  deformations of external kinematics \emph{uniquely} fixes the
  loop integrands in a number of non-trivial cases. In the integrand
  construction, all scaling conditions are homogeneous. Therefore, the only
  required information about the amplitude is its vanishing at particular
  points in momentum space. This homogeneous construction gives indirect
  evidence for a new geometric picture for graviton amplitudes similar to the
  one found for planar $\N=4$ super Yang-Mills theory. We also show how the
  behavior at infinity is related to the scaling of tree-level amplitudes under
  certain multi-line chiral shifts which can be used to construct new recursion
relations.}
\preprint{\begin{flushright} UUITP-38/19 \end{flushright}}

\begin{document}

\maketitle

\vspace{-1cm}
%=========================================================================
\section{Introduction}
\vspace{-.3cm}
%=========================================================================
%
The ultraviolet behavior of gravity scattering amplitudes has been of
great interest for several decades
\cite{tHooft:1974toh,Goroff:1985sz,Goroff:1985th,vandeVen:1991gw,Bern:2015xsa,Bern:2017puu}. Because
of the dimensionful coupling constant, perturbative gravity is expected to
develop ultraviolet (UV) divergences signaling the need for a UV
completion. Indeed, it was found a long time ago that scattering
amplitudes in Einstein gravity are UV divergent starting at two loops
\cite{Goroff:1985sz,Goroff:1985th,vandeVen:1991gw}. One well-known
mechanism to improve and tame the UV behavior of a theory is to
introduce supersymmetry which enforces certain cancelations of
divergences in loop diagrams due to superpartners running in the
loop. This famously leads to the cancelation of quadratic corrections
to the Higgs mass but naively it can not solve the problem in gravity
where power-counting would eventually win over any amount of
supersymmetry.

This expectation is related to the standard picture where UV divergences
of scattering amplitudes are closely linked to the appearance of
counterterms which satisfy all symmetry requirements of a given
theory. In this context, the existence of an $R^3$ counterterm in pure gravity is 
linked to the observed two-loop divergence. In contrast, supersymmetry
forbids the $R^3$ term and increases the loop order at which the
amplitude might diverge. For $\N=8$ supergravity
\cite{Cremmer:1978km,Cremmer:1978ds,Cremmer:1979up}, the allowed
counterterm consistent with all known symmetries of the theory has the
form $D^8R^4$ and implies a seven-loop divergence in four dimensions
\cite{Brodel:2009hu,Green:2010sp,Bossard:2010bd,Beisert:2010jx,Vanhove:2010nf,Bjornsson:2010wm,
Bjornsson:2010wu,Bossard:2011tq,Elvang:2010kc}. While
there is an ongoing debate whether or not this is indeed the case, indirect
evidence for the validity of the counterterm was given in
\cite{Bern:2018jmv} by calculating the five-loop UV divergence in the
critical dimension, $D_c = 24/5$, implying that the standard argument
holds \cite{Bjornsson:2010wm}\footnote{Note that perturbative finiteness of $\N=8$ SUGRA does not imply UV completeness \cite{Green:2007zzb}.}.

On the other hand, recent results for $\N<8$ supergravity \cite{Cremmer:1977tt} amplitudes 
suggest that our understanding of the relation between symmetries of
gravity theories and their UV structure is not yet satisfactory
\cite{Marcus:1985yy,Bern:2012cd,Bern:2012gh,Bossard:2012xs,Bern:2013qca,Kallosh:2016xnm,Freedman:2017zgq,Bern:2014sna,Bern:2017lpv,Bern:2017tuc,Bern:2017rjw,Bern:2019isl}. 
Perhaps this is due to our incomplete grasp of
supersymmetry itself and the lack of an off-shell superspace for
higher amount of supersymmetry. However, if some of the amplitudes' 
observed properties can not be explained by supersymmetry or duality
symmetries, it might point to new hidden symmetries or novel unexpected 
features of gravity.

For the time being, we would like to set aside the question of ultraviolet divergences in supergravity amplitudes. The aim of this paper, and more generally of the program initiated in
\cite{Bern:2017lpv,Herrmann:2018dja}, is to use instead the gravity \emph{loop integrand}
as probe to explore the UV physics of gravity amplitudes, ask basic
questions about analytic properties of gravitational scattering
amplitudes, and eventually connect them to geometric ideas such as the
Amplituhedron \cite{Arkani-Hamed:2013jha,Arkani-Hamed:2013kca,Arkani-Hamed:2017vfh,Damgaard:2019ztj} for
planar $\N=4$ super Yang-Mills (SYM) theory and other positive geometries \cite{Arkani-Hamed:2013jha,Arkani-Hamed:2017vfh,Arkani-Hamed:2018rsk,Arkani-Hamed:2017mur,Arkani-Hamed:2017tmz,He:2018okq}. In this approach, we
consider the ultraviolet region of amplitudes\footnote{We often use ``amplitudes'' synonymously with integrands of scattering amplitudes that still require integration over loop momenta.} as a broader concept.
The UV properties are not just a binary statement about the presence
or absence of divergences after integration, but more about the
behavior of the S-matrix at infinite loop momenta. While unitarity
implies the factorization of the S-matrix on infrared (IR) poles (at finite loop
momenta), an analogous statement is not known for UV poles at infinite
momenta --we denote those as \emph{poles at infinity}.

Naively, power-counting predicts the degree of the pole at infinity for a given theory and
should be manifest term-by-term in the expansion of amplitudes
in a basis of Feynman integrals. This picture also acts
behind the scene of most counterterm analyses, including the one for $\N=8$ supergravity.  
Whenever we can identify a divergent
integral in the expansion of the amplitude, we expect that this in turn
reflects the divergence of the full amplitude. Any possible UV
cancelations between terms that are \emph{not} a consequence of gauge
invariance or the known symmetries are therefore \emph{unexpected} and
directly point to some new property of the theory.

In \cite{Herrmann:2018dja}, two of the authors pointed out that there are
indeed cancelations which do improve the behavior of the loop integrand at
infinity in comparison to the UV scaling of individual terms. While we observed this phenomenon in
some isolated cases, in the present paper, we gather more comprehensive evidence and
provide new results in this direction. Very importantly, we are going to show
that the improved UV behavior of integrands is present only in $D=4$ due to vanishing Gram determinants.
This observation also explains the negative result in \cite{Bern:2017lpv} and suggests that
there are special features of four-dimensional gravity amplitudes still to be
discovered.

Furthermore, we demonstrate that the improved scaling at infinity is a 
powerful constraint in the construction of supergravity
amplitudes: in the generalized unitarity framework
\cite{Bern:1994cg,Bern:1997sc,Britto:2004nc,Bern:2007ct} it can be
combined with the scaling of tree-level amplitudes under BCFW
\cite{Britto:2004ap,Britto:2005fq} deformations to fix loop amplitudes
completely. All scaling constraints are \emph{homogeneous conditions},
i.e. we do not match the amplitude functionally on cuts but rather
demand that the unitarity based ansatz for the amplitude vanishes at
certain points at infinity. The fact that homogeneous conditions are
sufficient to uniquely fix gravity amplitudes also suggests a possible
connection to the Amplituhedron geometry, in analogy to the discussion
for $\N=4$ SYM theory beyond the planar limit \cite{Bern:2015ple}.

It is important to note that our discussion concerns the cuts of loop
integrands. Based on unitarity, these cuts are given by products of
tree-level amplitudes. Therefore, the behavior of loop integrands at
infinite loop momenta is linked to large momentum shifts of trees. 
It has been known for a while that graviton
tree-level amplitudes have a surprisingly tame large $z$ behavior for BCFW
shifts \cite{Bedford:2005yy,Cachazo:2005ca,Benincasa:2007qj} despite
the naive power-counting expectations. This feature of gravity trees has been linked to improved
UV properties of one-loop amplitudes in e.g.~\cite{Bern:2007xj}. 
Here, we show that there are more general shifts of tree-level
amplitudes with similar properties that can be used to reconstruct all
graviton tree-level amplitudes.

The remainder of this work is structured as follows: In Section \ref{sec:integrands_cuts}, 
we summarize salient features of the unitarity method and explain 
how basic UV properties of the diagrammatic expansion of amplitudes 
can be extracted from maximal cuts. In Subsections \ref{subsec:poles_infty} and 
\ref{subsec:cancelations}, we concretize the notion of a pole at infinity 
and potential cancelations thereof in the context of cuts. In section \ref{sec:scaling},
we present one of the main results of our work. We analyze the scaling of 
multi-particle unitarity cuts for Yang-Mills and gravity in both general $D$ and $D=4$.
We find a surprising drop in the large momentum scaling in 
gravity when going to $D=4$ which is attributed to the vanishing of a certain Gram 
determinant. In section \ref{sec:loop_construction} we lay out our second new result. 
We show, that the large momentum scaling behavior together with a few other homogeneous 
constraints are sufficient to uniquely fix the $\N=8$ supergravity amplitude through three-loops and 
four external particles. In section \ref{sec:tree_recursion}, we attempt to understand some 
of the observed large momentum scaling improvements of gravity unitarity cuts in terms 
of properties of tree-level amplitudes under generalized shifts. We point out that under
certain conditions, these new shifts lead to novel recursion relations of gravity tree-level amplitudes.
We close in section \ref{sec:conclusion} with some conclusion and an outlook to future work.

%=========================================================================
\section{Integrands and cuts}
\label{sec:integrands_cuts}
%=========================================================================

The textbook formulation for the perturbative S-matrix is based on the
expansion of scattering amplitudes in terms of Feynman
diagrams. Higher order corrections in the perturbative series are
encoded in loop amplitudes. For the $L$-loop $n$-particle amplitude in
$D$ spacetime dimensions we can write,
\begin{equation}
\label{eq:FD_expansion}
\A^{L-\text{loop}}_n = \sum_{\text{FD}} \int d^D\ell_1\d^D\ell_2\dots d^D\ell_L\, \I_{n}^{\text{FD}}\,,
\end{equation}
where $\I_{n}^{\text{FD}}$ is a rational function of external momenta,
loop momenta, polarization states, and possibly gauge theory data. The
only poles in $\I_{n}^{\text{FD}}$ come from Feynman propagators and have
the form $1/P^2$, where $P$ schematically represents a combination
of external and loop momenta. Individual Feynman diagrams are not
gauge invariant while the full amplitude $\A^{L-\text{loop}}_n$ is. We
can decompose all Feynman diagrams into a basis of independent
integrands (scalar integrals). The resulting decomposition of the
amplitude is a linear combination of these basis elements with gauge
invariant coefficients $c_k$,
\begin{equation}
\label{eq:diag_expansion}
\A^{L-\text{loop}}_n = \sum_k c_k\,I_k \quad \mbox{where} \quad I_k = \int d^D\ell_1\d^D\ell_2\dots d^D\ell_L\, \I_k\,.
\end{equation}
Searching for bases of loop integrands $\I_k$ is a very active area of
research and many efficient methods have been developed in recent
years to perform these calculations to higher multiplicities and
higher loops in wide range of QFTs
\cite{Passarino:1978jh,Ossola:2006us,Mastrolia:2010nb,Ita:2015tya,Bourjaily:2017wjl}.

In the planar limit we can exchange the sum and the integration symbol
and define the loop integrand $\I^{L-\text{loop}}_n$ as the sum of all
contributing pieces prior to integration
\begin{equation}
\A^{L-\text{loop}}_n = \int d^D\ell_1d^D\ell_2\dots d^D\ell_L\,\, \I_{n}^{L-\text{loop}}\,.
\end{equation}
It has been demonstrated in a number of cases that the loop integrand
is not just an intermediate object in the calculation but rather it
exhibits some remarkable properties deserving of an independent raison
d'\^etre. Prominent examples include new methods for constructing the
planar $\N=4$ SYM integrand using loop recursion relations
\cite{ArkaniHamed:2010kv}, the connection to on-shell diagrams and
Grassmannian \cite{ArkaniHamed:2012nw}, and the complete reformulation
using the geometric Amplituhedron picture 
\cite{Arkani-Hamed:2013jha,Arkani-Hamed:2013kca,Arkani-Hamed:2017vfh,Damgaard:2019ztj}. In contrast, there
are a number of approaches advocating to calculate amplitudes directly without
ever discussing integrands. These are based on bootstrap ideas of writing down
appropriate function spaces for scattering amplitudes and imposing
physical conditions to uniquely extract the scattering amplitudes, 
see e.g.~\cite{Goncharov:2010jf,Drummond:2014ffa,Dixon:2011nj,Dixon:2011pw,Caron-Huot:2019vjl}. 

%=========================================================================
\subsection{Perturbative unitarity}
\label{subsec:pert_unitarity}
%=========================================================================

Beyond the planar limit, the loop integrand can not be defined in the
same way due to the lack of global variables\footnote{See \cite{Ben-Israel:2018ckc,Tourkine:2019ukp} for recent progress in that direction.}. 
Instead, we have to adhere to
the diagrammatic expansion in Eq.~(\ref{eq:diag_expansion}). However, the
loop integrand is still a very important concept which underlies the
success of unitarity methods. Perturbative unitarity implies that the
loop amplitude must factorize into lower-loop amplitudes when evaluated
on \emph{cuts}. In the most basic unitarity cut, two
inverse propagators are set on-shell, $\ell^2=(\ell+Q)^2=0$ and
the amplitude factorizes into two pieces\footnote{\label{fn:drop_dlips}In the following, we will drop the integration over the Lorentz invariant phase space $d\text{LIPS}_\ell$.},
\begin{align}
\begin{split}
\label{eq:cut_L_loop}
\underset{\begin{subarray}{c}
   \ell^2{=}0\\
  (\ell{+}Q)^2{=}0
  \end{subarray}}{\rm Cut}\bigg[\A_n^{L-\text{loop}}\bigg]  & =  \underset{\begin{subarray}{c}
  L{=}L_1{+}L_2{+}1 \\
  \rm  states
  \end{subarray}}{\sum}  \int d{\rm LIPS}_{\ell} \,\A_{n_1+2}^{L_1-\text{loop}} \times \A_{n_2+2}^{L_2-\text{loop}} \\
  \underset{\begin{subarray}{c}
   \ell^2{=}0\\
  (\ell{+}Q)^2{=}0
  \end{subarray}}{\rm Cut} \left[\raisebox{-28pt}{\includegraphics[scale=.5]{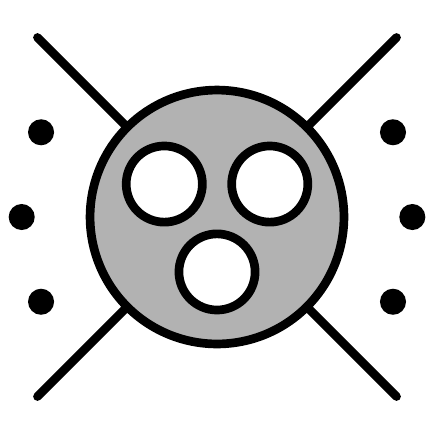}} \right] &= \underset{\begin{subarray}{c}
  L{=}L_1{+}L_2{+}1 \\
  \rm  states
  \end{subarray}}{\sum}  \int d{\rm LIPS}_{\ell}\,  \raisebox{-35pt}{\includegraphics[scale=.5]{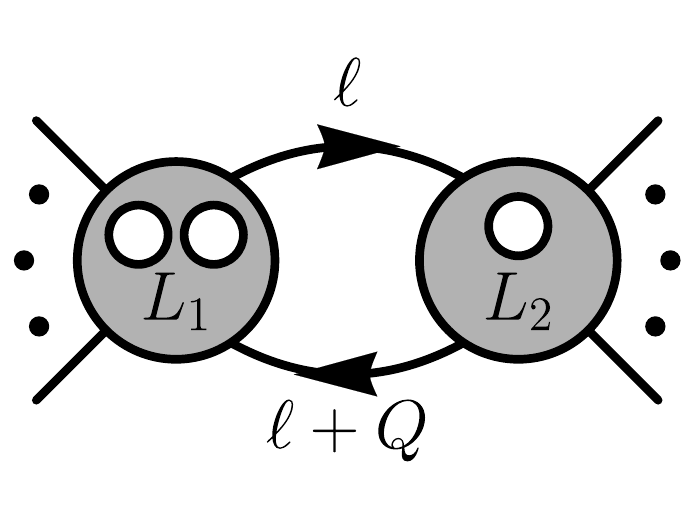}}
\end{split}  
\end{align}
where the sum is over the distribution of loop orders $L_1,L_2$ as
well as the allowed on-shell states exchanged in the cut. The
distribution of external legs $n_1,n_2$ of the subamplitudes have to
be consistent with the cut channel $Q$ and are related to the number
of external states $n$ via $n=n_1{+}n_2$. The unitarity cut
(\ref{eq:cut_L_loop}), and the basic tree-level factorization
\begin{align}
\begin{split}
 \underset{Q^2{=}0}{\rm Cut} \left[ \A^{\text{tree}}_n \right] 
 	&= \sum_{\text{states}} \A^{\text{tree}}_{n_1+1} \times \A^{\text{tree}}_{n_2+1} \\[-15pt]
 \underset{Q^2{=}0}{\rm Cut}\left[ \raisebox{-22pt}{\includegraphics[scale=.4]{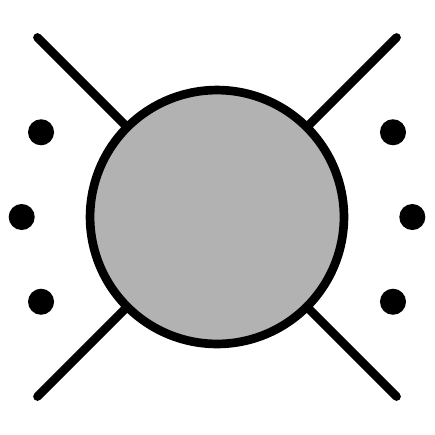}} \right] 
	&= \underset{  \rm  states}{\sum} \raisebox{-35pt}{\includegraphics[scale=.5]{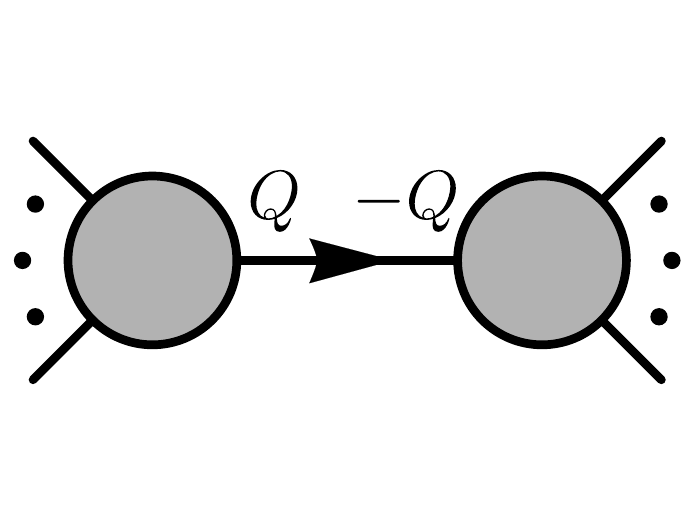}}
\end{split}
\end{align}
can be iterated to give rise to \emph{generalized unitarity} \cite{Bern:1994cg,Bern:1997sc,Britto:2004nc,Bern:2007ct}. 
In this setup, we can set to zero any number of propagators and the loop
amplitude factorizes correspondingly\footnote{In massless theories in $D=4$, the three particle amplitudes are special and completely fixed by Lorentz invariance. Momentum conservation and the on-shell conditions allow for MHV (blue vertex) and $\bar{\text{MHV}}$ (white vertex) amplitudes, see e.g.~\cite{ArkaniHamed:2012nw,Herrmann:2016qea} for more details.}. 

\begin{equation}
  \raisebox{-20pt}{\includegraphics[scale=.5]{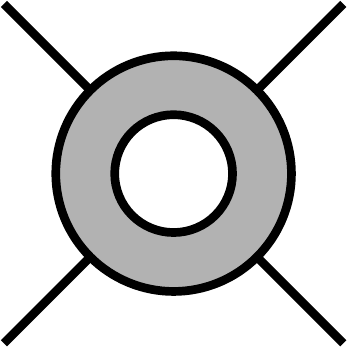}}
  \overset{\rm Res}{\longrightarrow}  
  \raisebox{-20pt}{\includegraphics[scale=.5]{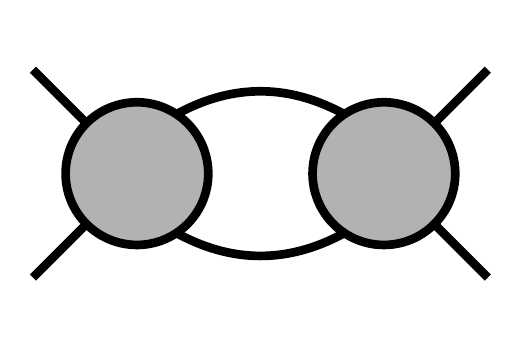}} 
  \overset{\rm Res}{\longrightarrow}  
  \raisebox{-20pt}{\includegraphics[scale=.5]{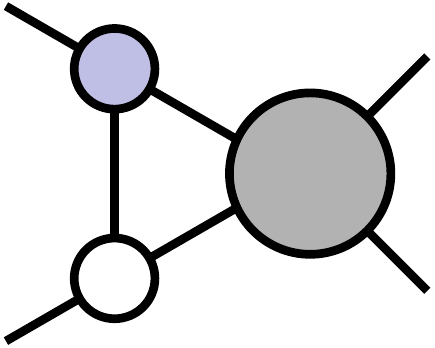}}
  \overset{\rm Res}{\longrightarrow}  
  \raisebox{-20pt}{\includegraphics[scale=.5]{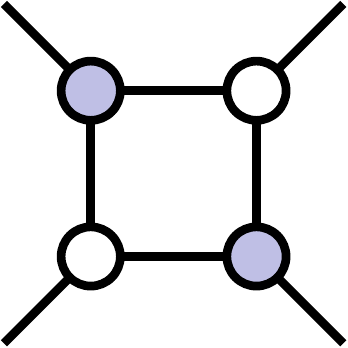}} 
\vspace{10pt}
\end{equation}

This can be viewed as modifying
the contour of integration to encircle poles (changing $R^{3,1}$ to
involve $S^1$ around the poles), or equivalently, as taking residues
of the loop integrand (see e.g.~\cite{Cachazo:2008vp}). While the loop
integrand $\I_{n}^{L-\text{loop}}$ is not a unique
rational function beyond the planar limit (due to the aforementioned lack of global variables),
the unitarity cuts are still well-defined. In particular, the
uniqueness and associated label problem is completely avoided if we
consider situations where each loop is cut at least once and the residue
is a product of tree-level amplitudes, as in~(\ref{fig:MUC}) and (\ref{fig:muc_diags}). 

The labels of the basis integrands $\I_k$ contributing to the expansion of the cut amplitude are
unambiguously linked to on-shell legs in tree-level
amplitudes. Importantly, we do not need to know the full amplitude
beforehand in order to calculate unitarity cuts (As explained above,
cuts are gauge invariant objects given by products of tree-level
amplitudes.). There is no issue about basis choices, ambiguity of
labelings or total derivatives etc. Knowing high multiplicity
tree-level amplitudes suffices to calculate very high loop cuts, even
if we do not have direct access to (uncut) amplitudes. 

The unitarity cuts provide a considerable amount of information about
the original loop integrand and indirectly about the loop
amplitude. In \emph{cut-constructible} theories \cite{Bern:1996je}
this information is complete, i.e. knowing all four-dimensional cuts allows us to
uniquely reconstruct the loop integrand. In other cases, we have
to include extra information. This can include soft or collinear
limits, or knowing $D$-dimensional cuts \cite{Anastasiou:2006jv}. Therefore,
it is fair to say that cuts indeed specify the loop integrand
uniquely, despite its explicit construction might be laborious and not
practical for higher loops, e.g. due to the missing knowledge of
the integrand basis.

The connection between properties of the loop integrand, its cuts, and
the final amplitude is a very difficult question, but in certain cases
we do have a partial or even complete understanding. In particular,
the IR divergences of the amplitude come from very well-known regions
of the loop integration, and are captured by soft and collinear
cuts. In other words, any integrand which vanishes on these cuts must
be IR finite and vice versa. Another peculiar feature is the uniform
transcendentality property of certain integrals and $\N=4$ SYM
amplitudes: the integrals evaluate to polylogarithms of uniform
degree. (For sufficiently complicated amplitudes and integrals, the
space of polylogarithmic functions is insufficient, see
e.g. \cite{CaronHuot:2012ab,Bourjaily:2017bsb,Bourjaily:2018ycu,Bourjaily:2018yfy,Broedel:2018qkq}). This is closely related to logarithmic ($\dlog$) singularities of the loop integrand and 
underlies much of the geometric story behind on-shell diagrams, the
positive Grassmannian and the Amplituhedron. More practically, all
these properties have been used to construct special integrands
\cite{Arkani-Hamed:2014via,Bern:2014kca,Bern:2015ple} that are relevant for deriving
differential equations for families of Feynman integrals in canonical
form \cite{Henn:2013pwa,Henn:2014qga,Chicherin:2018wes}. On more general grounds, the
cuts of loop integrands are related to the branch cuts of final
amplitudes (for recent work in this direction for Feynman integrals,
see e.g.~\cite{Abreu:2017ptx,Abreu:2017enx}), despite a detailed link
is not yet completely understood.

%=========================================================================
\subsection{Cuts and UV}
\label{subsec:cuts_uv}
%=========================================================================

In the context of cuts, it is natural to ask how the UV behavior of amplitudes is encoded in
loop integrands. On one hand, this has a simple answer: the UV
divergences come from regions of large loop momenta. It is also
relatively straightforward to determine the critical dimension
$D_c$,~i.e.~the spacetime dimension where the first logarithmic
divergence appears. This is done by rescaling the loop variables
$\ell_k\rightarrow t\,\widetilde{\ell}_k$ and asking for what value of
$D_c$ the integrand scales asymptotically like $dt/t$ as
$t \to \infty$. As a trivial example, consider the scalar bubble
integral at one loop. The $\ell\rightarrow t\,\widetilde{\ell}$
rescaling effectively corresponds to introducing a radial coordinate
$t$. Transforming the measure
$d^D\ell \to t^{D-1} dt \,d^{D-1}\widetilde{\ell}$ and neglecting the
angular coordinates $\widetilde{\ell}$, we find
\begin{equation}
I_{b} = 
\raisebox{-22pt}{\includegraphics[scale=.5]{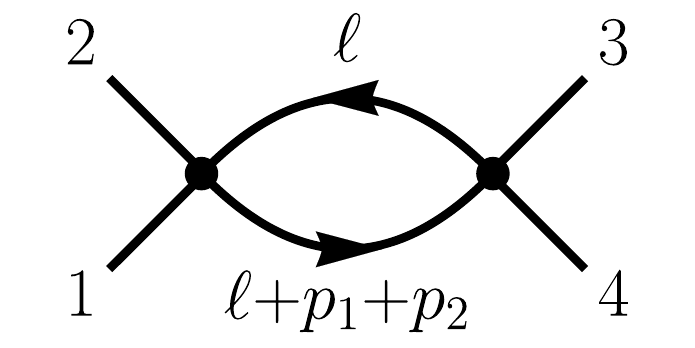}}
= \int \frac{d^D\ell}{\ell^2(\ell{+}p_1{+}p_2)^2} \xrightarrow[t\rightarrow\infty]{} \int \frac{dt}{t^{5-D}}\,,
\end{equation}
that the critical dimension for the bubble is $D_c=4$. Said
differently, fixing the spacetime dimension to $D=4$, the bubble
integral is logarithmically divergent, while scalar triangles and
boxes are UV finite. This scaling analysis is exactly what is
traditionally understood as power-counting loop momenta, and unless the
remaining integral vanishes for auxiliary reasons, we learn everything
about the presence of UV divergences of an integral from the large $t$ behavior.

Performing a similar analysis for the full amplitude is a bit more
subtle. First, if we expand the amplitude in terms of
Feynman diagrams (\ref{eq:FD_expansion}) there could be cancelations
between different diagrams as a consequence of gauge invariance. In
order to account for such cancelations, it is preferential to express the amplitude in terms
of basis integrals with gauge invariant coefficients
(\ref{eq:diag_expansion}).  In this case, barring any further
surprises, it is expected that the UV behavior of the amplitude is
given by the worst behaved integral.

This begs the immediate question: Is there an invariant way to
determine the \emph{minimal}\footnote{
 Roughly, ``minimal power-counting'' denotes numerator polynomials 
 with the lowest possible degree in the loop variables $\ell_i$. 
 For a  detailed definition and various subtleties, see e.g.~\cite{Bourjaily:2020qca}. 
 Note that one can always write a basis of integrands with higher power-counting that contains
 the minimal power-counting basis as a subspace. Superficially boosting
 the power-counting this way is not what we mean here.} power-counting
of an integral?  The answer is that power-counting of individual
integrals is dictated by the method of \emph{maximal cuts} and thus by well
defined, gauge invariant data of the theory itself. We consider a cut
of the amplitude where the maximal number of propagators are set
on-shell. This maximal cut singles out one contributing basis integral
and its numerator must have the appropriate form to match the cut
calculated as the product of tree-level amplitudes. Therefore,
numerators for integrals with the maximal number of propagators (also
called parent integrals) are fixed by maximal cuts. We can always add
contact terms (shrinking propagators of parent integral) to a parent
diagram and rotate the basis but this does not change the
power-counting of the irreducible piece which is uniquely associated to
the parent diagram and is required to match the maximal cut functionally.

One particular example to have in mind is an integral which
contributes to the four-particle $\N=8$ supergravity amplitude and
will play a role in our later discussion. The maximal cut corresponding to this integral is
\begin{equation}
\label{eq:max_cut_eg}
  {\rm MaxCut}\left[\A^L_4\right] = \raisebox{-53pt}{\includegraphics[scale=.5]{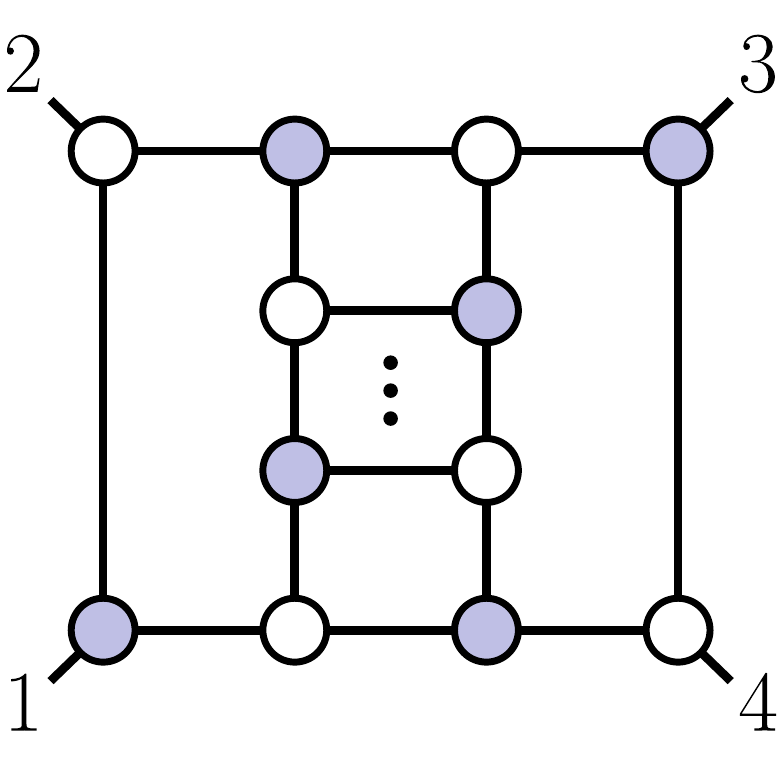}} =  \raisebox{-53pt}{\includegraphics[scale=.5]{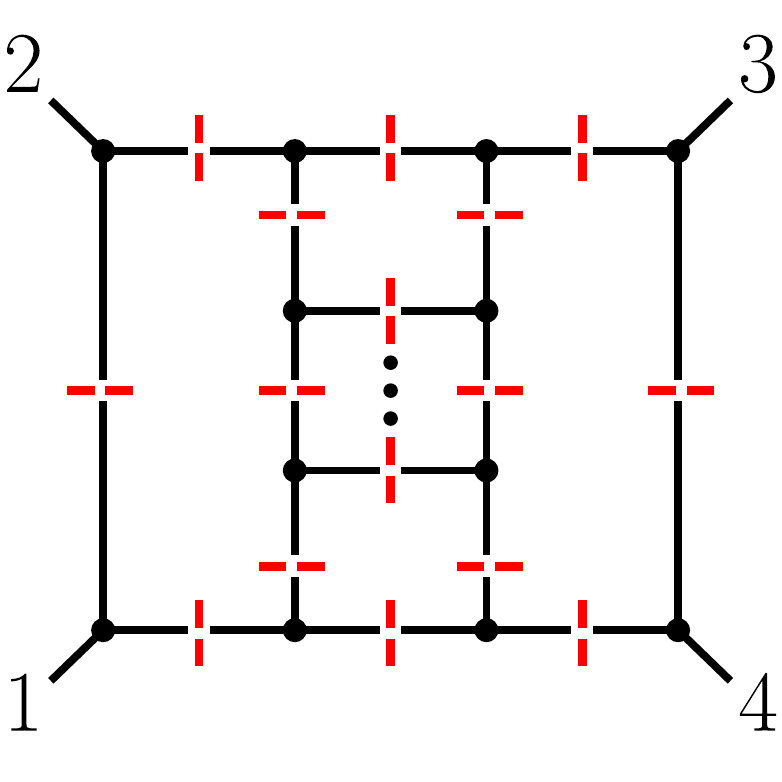}}\, N(\ell_i,p_j)\,.
\end{equation}
Matching the field theory cut of $\N=8$ SUGRA on the left hand side of (\ref{eq:max_cut_eg}) requires 
the numerator of the local diagram to be $N(\ell_i,p_j)\!=\!(\ell_1\!\cdot\! \ell_2)^{2(L{-}3)}$ modulo terms of
lower power-counting in the $\ell_i$, or terms which vanish on this maximal cut (contact terms).
\begin{equation}
  \raisebox{-53pt}{\includegraphics[scale=.5]{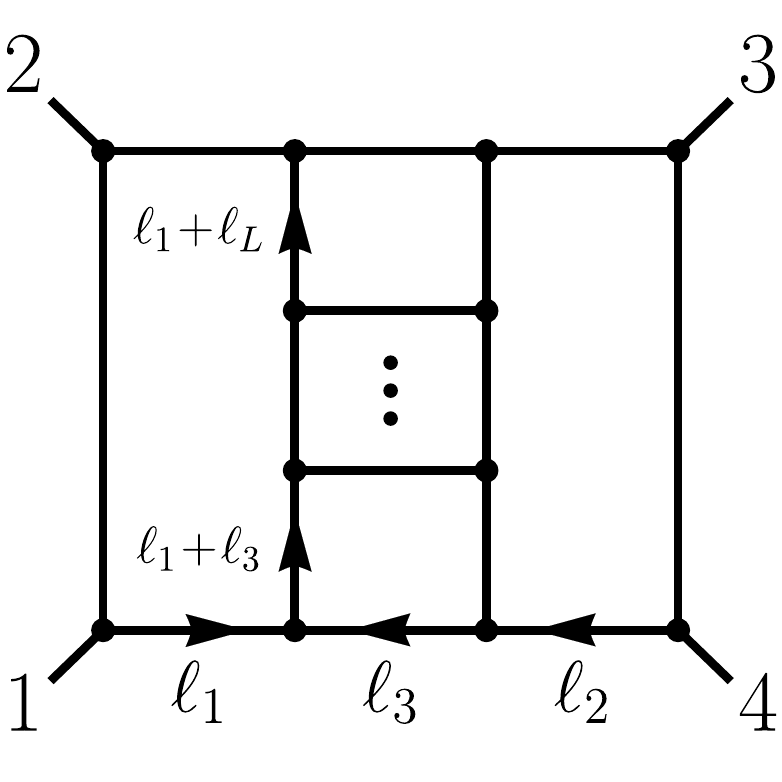}} \quad N(\ell_i,p_j) = \begin{cases}
    f_{\rm YM} \times (\ell_1\cdot \ell_2)^{L-3}, \quad &\N=4 \,\,{\rm SYM}\\
    f_{\rm GR} \times (\ell_1\cdot \ell_2)^{2(L-3)}, \quad & \N=8\,\, {\rm SUGRA}
\label{fig:bad_scaling_diag}
\end{cases}
\end{equation}
In fact, (\ref{fig:bad_scaling_diag}) is a representative of the
worst behaved diagram relevant for supergravity amplitudes in the
UV. Continuing this line of logic, we see that this integral is
divergent for $L\geq7$ in four dimensions which suggests the presence of
the $D^8R^4$ counterterm in $\N=8$ supergravity. If extrapolating
the UV divergence of the full amplitude from the worst behaved local integral 
were legitimate, we would conclude that the amplitude indeed
diverges starting at seven loops. Note that the power-counting of
$\N=4$ SYM is such that all diagrams stay UV finite to any loop order
in $D=4$.

There is an obvious caveat in the extrapolation argument: it is possible for 
UV divergences to cancel between various seven-loop diagrams,
making the final amplitude UV finite (in $D=4$). This would then
result in a zero coefficient for the $D^8R^4$ counterterm. A direct
seven-loop calculation is not within current reach, but analogous
$\N<8$ calculations revealed that at lower loops there indeed occur
\emph{enhanced} cancelations of UV divergences between various terms
making the result surprisingly finite
\cite{Bern:2012cd,Bern:2014sna,Bern:2012gh,Bern:2017lpv}. 
(A detailed discussion of the status of UV divergences in non-maximal 
supergravity theories, and various string 
and symmetry based analyses are beyond the scope of this work and a review 
can be found in the introduction sections of most of the references cited here.)
On the other hand,
the direct $\N=8$ supergravity calculation for $L=5$ in the critical
(fractional) dimension $D_c=24/5$ showed that there were no cancelations of
this sort and the naive power-counting extrapolation was indeed the correct one
\cite{Bern:2018jmv}. Conservatively, this seems to seal the fate of the $D^8R^4$ 
counterterm with an expected UV divergence at seven loops, assuming there is 
nothing special about $D=4$ compared to the general $D$-dimensional gravity 
amplitudes. While we can not claim anything concrete about UV divergences 
in this work, we will show that four-dimensional gravity loop 
integrands indeed behave in a surprisingly good way.

%=========================================================================
\subsection{Poles at infinity}
\label{subsec:poles_infty}
%=========================================================================

Instead of a direct integration approach which faces technical
challenges when attempting to go to seven loops, we take a different
path to explore the physics of the UV structure of gravity. In
particular, we focus on poles at infinity in the loop integrand
evaluated on unitarity cuts. On one hand, studying the behavior of
cuts does not directly tell us much about the UV divergences of
the full amplitude as performing cuts effectively changes the
contour of integration (see discussion in
subsec.~\ref{subsec:pert_unitarity}). On the other hand, we gain
access to a richer set of statements about the behavior of the loop
integrand at infinite loop momenta, beyond a binary statement about
the presence or absence of a UV divergence. In particular, we are
interested in the broader question of how physical principles
constrain the behavior of the loop amplitude at infinity. As
summarized in the beginning of Sec.~\ref{sec:integrands_cuts}, we know that unitarity
dictates that the loop integrand factorizes when evaluated on the
propagator poles. These factorization poles are in the IR (at finite
momentum), but no analogous statements are known about the poles at
infinite loop or external momenta. The behavior at infinity is also
closely related to symmetries. In planar $\N=4$ SYM for example, the
(complete) absence of poles at infinity is a direct consequence of
dual conformal symmetry \cite{Drummond:2007au,Drummond:2006rz}.

The aforementioned UV scaling $\ell \rightarrow t\,\widetilde{\ell}$
determines the presence (and degree) of UV divergences but probes
infinity in a generic direction $\widetilde{\ell}$. As we will see
later, there are special directions $\ell\rightarrow t\ell^\ast$ with
$t\rightarrow\infty$ where the naive (power-counting) expectation does
not work and the pole at infinity is absent (or has lower
degree). These directions naturally appear on cut surfaces where the
loop momentum gets partially fixed by on-shell conditions. Starting
from the cut surface we subsequently send the loop momenta to infinity
respecting the on-shell conditions.
\begin{equation}
{\rm Cut} \Bigg[ \raisebox{-28pt}{\includegraphics[scale=.5]{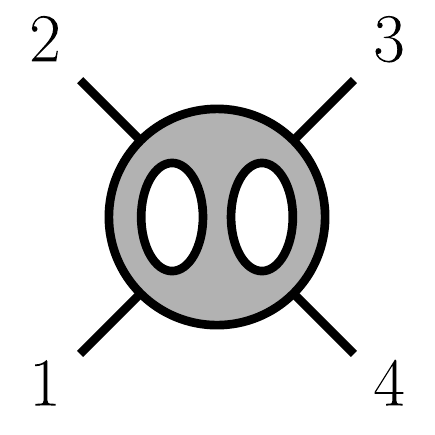}} \Bigg] = \underset{  \rm  states}{\sum}  
\raisebox{-29pt}{\includegraphics[scale=.5]{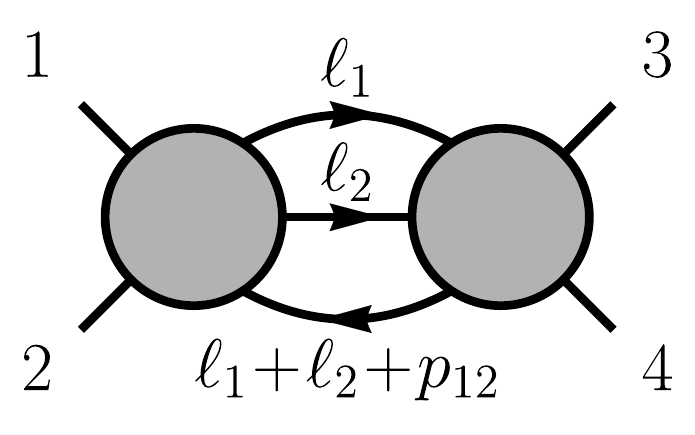}} \longrightarrow \quad \begin{matrix}
  \ell_1^* = t \lambda_{\ell_1}\tilde\lambda_{\ell_1}\\
  \ell_2^* = t \lambda_{\ell_2}\tilde\lambda_{\ell_2} 
\end{matrix}
\end{equation}
In fact, the necessity to first cut and then send the loop momentum to
infinity is not optional and is forced on us if we want to discuss the
behavior of the full loop integrand, not just individual basis
integrals. This is because approaching the poles at infinity directly
suffers from the same labeling problem described in
subsec.~\ref{subsec:pert_unitarity}: without cutting, $\ell$ means
different things in different diagrams, and therefore asking for a
global meaning of $\ell\rightarrow\infty$ is ill-defined.
\begin{figure}[htb]
  \centering
\raisebox{-35pt}{
\includegraphics[scale=.5]{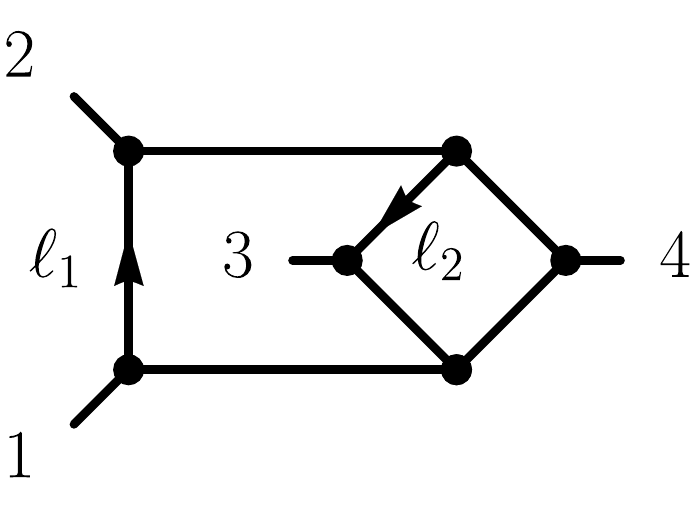}}
\quad \text{What are $\ell_1$ and $\ell_2$?} \quad
\raisebox{-35pt}{
\includegraphics[scale=.5]{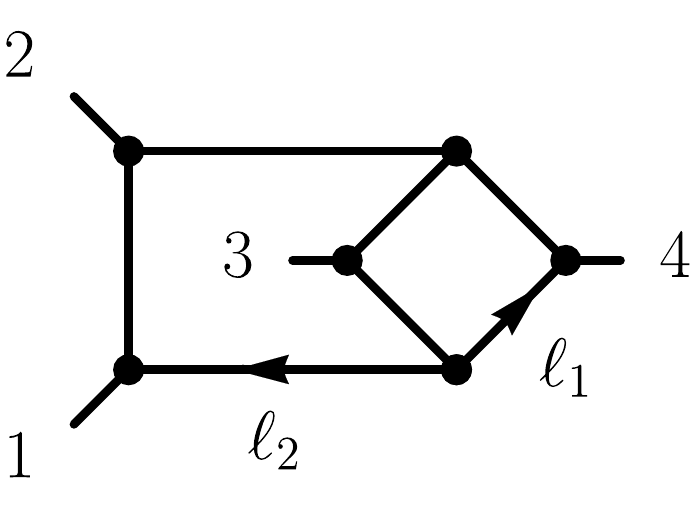}}
\caption{Ambiguity in labeling loop momenta in a given contribution to the integrand.}
\end{figure}

To be able to approach the UV limit in a well-defined manner we
therefore have to first cut a certain number of propagators. In
particular, we have to cut all loop momenta, at least in the minimal
way, to factorize the loop integrand as a product of tree-level
amplitudes. Then we can scale these cut loop momenta to infinity in
various ways and ask how the integrand behaves under these scalings.

%=========================================================================
\subsection{Cancelations}
\label{subsec:cancelations}
%=========================================================================

As we discussed before, one can compute the cut function as a product
of tree amplitudes and perform the scaling limits explicitly without
knowing the full integrand  in the
first place (by full integrand, we mean the knowledge
of all coefficients $c_k$ in Eq.~(\ref{eq:diag_expansion})). 
However, it is still very useful to compare a particular
behavior at infinity of the (cut) loop integrand with the behavior of
basis integrands ($\I_k$ in Eq.~(\ref{eq:diag_expansion})) which
contribute to the amplitude. The conservative expectation is that the
scaling of the loop integrand on a particular pole at infinity is
dictated by the basis integrands with the worst UV behavior (highest
degree pole in the large $t$ limit). In the extreme case of maximal
cuts this is indeed the case: only one basis integrand contributes and
the behavior of the loop integrand is given by this term. In fact,
this was used in subsection \ref{subsec:cuts_uv} to determine the
power-counting.

If we cut fewer propagators, more basis integrands contribute, and
there is a chance for cancelations. We initiated this work in
\cite{Herrmann:2018dja} for various cuts in $D=4$ and indeed found
such cancelations where the loop integrand is better behaved at
infinity than individual terms. While this initial study was very
suggestive, it left some important questions unanswered: What is the
role of $D=4$ vs general $D$? Are the cancelations present only for
special cuts? What are implications for the final amplitude?

We will answer the first two questions in this paper, while the third
(most difficult) has to be relegated to future work. We know that the
complete absence of poles at infinity leads to a simpler structure of
integrated results.  However, it is not clear how the absence of a
particular pole at infinity is encoded in the final integrated answer.

We mainly focus on the most minimal cuts which specify unique
labels and therefore allow us to talk about poles at infinity for the
full (cut) loop integrand. From this perspective, the multi-particle 
unitarity cut is a prime representative,
\begin{align}
\begin{split}
\label{fig:MUC}
F(\ell_k,p_j) 
=  \raisebox{-35pt}{\includegraphics[scale=.5]{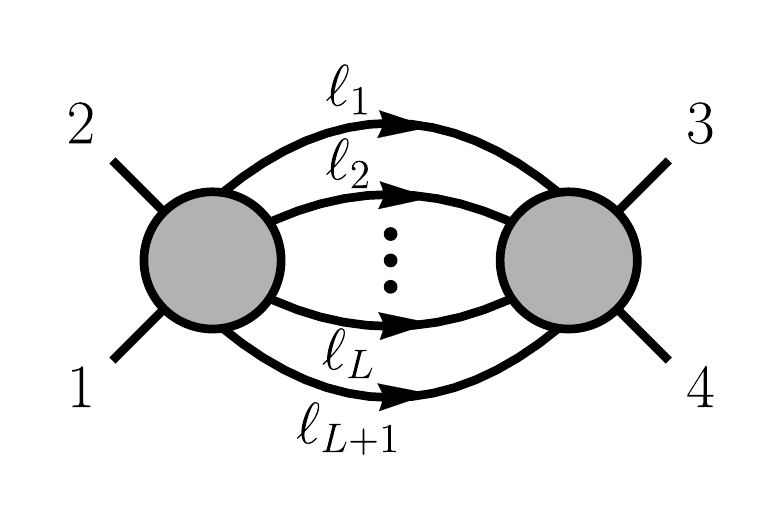}}  \hspace{-.3cm} 
= \underset{\begin{subarray}{c}
   \ell^2_k{=}0
  \end{subarray}}{\rm Cut} \!\left[\A_n^{L-\text{loop}}\right] 
=  \sum_{\text{states}} \A^{\text{tree}}_{2+L+1}\times \A^{\text{tree}}_{2+L+1}\,.
\end{split}
\end{align}
The residue of the loop amplitude on this cut is given by the product
of two tree-level amplitudes (integrated over the remaining phase
space [see footnote \ref{fn:drop_dlips}] and summed over the exchanged on-shell states). Our goal is to
study the behavior of this cut in the UV region where the on-shell
(cut) loop momenta $\ell_k$ approach infinity,
$\ell_1,\ldots,\ell_L,\ell_{L{+1}}\rightarrow\infty$, and compare the full cut to the
contributing basis integrands,
\begin{equation}
\label{fig:muc_diags}
\raisebox{-45pt}{
 \includegraphics[scale=.51]{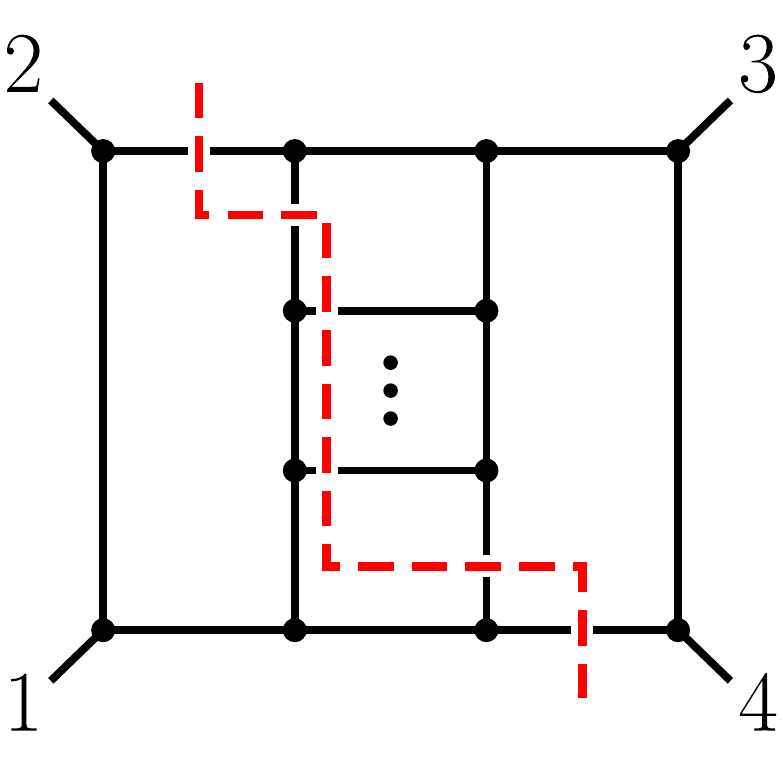}}
 \qquad
\raisebox{-43pt}{
 \includegraphics[scale=.49]{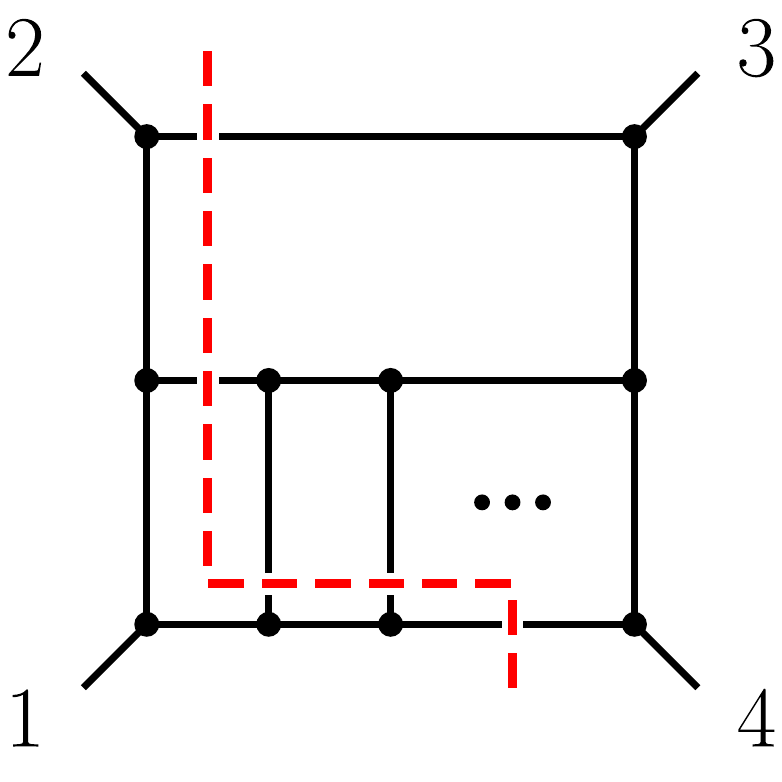}}
\end{equation}
The only comparable analysis was done for certain two-loop four-point
amplitudes \cite{Bern:2017lpv}, where it was concluded that the
$D$-dimensional amplitude has the same scaling as the contributing
integrals and no cancelations occur. In \cite{Herrmann:2018dja} the
analysis was repeated in $D=4$ finding an improved scaling at
infinity of the cut amplitude compared to individual integrals. 
This signals that cancelations indeed happen in $D=4$.

%=========================================================================
\section{Improved scaling at infinity, general $D$ vs. $D=4$}
\label{sec:scaling}
%=========================================================================

We first focus on the multi-particle unitarity cut illustrated in 
(\ref{fig:MUC}) for four-particle amplitudes in $D$ dimensions.
All internal propagators visible in the figure are cut and impose
$L+1$ on-shell conditions,
\begin{equation}
\label{eq:cut_legs_muc}
\ell_1^2=\ell_2^2=\dots=\ell_L^2 = \ell_{L+1}^2=0 \quad \mbox{where}
\quad
\sum_{k=1}^{L+1} \ell_k = -(p_1{+}p_2)\,.
\end{equation}
On support of these cuts, $F(\ell_k,p_j)$ is a $(D{-}1)L{-}1$ parametric
function of on-shell momenta $\ell_k$ which satisfy momentum
conservation as in Eq.~(\ref{eq:cut_legs_muc}). On this cut surface, there
are numerous options how to scale the on-shell momenta $\ell_k$ to
infinity. A very general way how to do this scaling is to perform a
shift
\begin{equation}
\label{eq:scal_deformation}
\ell_k \rightarrow \ell_k + t \,q_k\,,\quad \mbox{where} \quad (\ell_k\cdot q_k)=q_k^2=0\,,\,\,\,\mbox{and} \,\,\, \sum_k q_k = 0\,.
\end{equation}
The conditions imposed on the $q_k$ guarantee momentum
conservation and the on-shellness of the shifted momenta. Under this
shift we get another on-shell function $F(\ell_k,q_k,p_j,t)$ which now
depends not only on the original momenta $p_j$, $\ell_k$ but also the
shift parameters $q_k$ and $t$. We approach infinity by scaling
$t\rightarrow\infty$ keeping $q_k$ generic, and organize the result as
a series in $t$,
\begin{equation}
  \label{eq:cut_series}
\lim_{t\rightarrow\infty} F = t^m\,F_m + {\cal O}(t^{m-1})\,.
\end{equation}
We are interested in the parameter $m$ which controls the leading
behavior of the cut integrand at infinity. For general $q_k$, we
indeed find that the behavior of the $\N=8$ supergravity, as well as
the pure gravity loop integrand, is controlled by the worst behaved
local diagrams such as the one depicted in
Fig.~\ref{fig:bad_scaling_diag} for $L\geq4$. This is absolutely
expected as a drop in the exponent for general shift values $q_k$
would very likely indicate a decrease in power-counting and therefore,
an increase of the critical dimension for the UV divergence. However,
from the analysis of $\N=8$ as well as pure gravity amplitudes we know
that this can not be the case.

%=========================================================================
\subsection{Special shift in $D$ dimensions}
\label{subsec:special_shift}
%=========================================================================

We choose to further specialize our shift (\ref{eq:scal_deformation}) to the subspace defined by
\begin{equation}
\label{eq:spec_q_constraint}
(q_i\cdot q_j) = 0\qquad \mbox{for all $i,j$}\,,
\end{equation}
where the shifted propagators are all linear in $t$ for $t\rightarrow\infty$,
\begin{equation}
(\ell_i {+} \ell_j {+}Q)^2 \rightarrow (\ell_i {+} t\,q_i {+} \ell_j{+}t\,q_j {+} Q)^2 \sim {\cal O}(t)\,,
\end{equation}
since the quadratic terms in $t$ cancel. Performing the calculation explicitly
for $\N=8$ SUGRA and $\N=4$ SYM, we see that the loop integrand scales
like
\begin{align}
F_{\text{SUGRA}}\sim \frac{1}{t^4}\,,
\qquad 
F_{\text{SYM}} \sim \frac{1}{t^{L+2}}\,, 
\end{align}
which is in agreement with the scaling of the worst behaved diagrams,
and no cancelations occur. In fact, in order to perform these
$D$-dimensional scaling analyses, we analyzed the results constructed
in \cite{Bern:1998ug,Bern:2008pv,Bern:2009kd} and calculated the
scaling from these integrand representations rather than gluing
tree-level amplitudes together. The reason for doing so is to avoid
technical complications involved with higher multiplicity
$D$-dimensional tree-level amplitudes.
\begin{table}[h!]
\centering
\begin{tabular}{|c|c|c|c|}
 \hline
 		& $L=2$ 	& $L=3$ 	& $L=4$   \\
 \hline		
 SUGRA	& $t^{-4}$ & $t^{-4}$ & $t^{-4}$ \\
 \hline
 SYM	& $t^{-4}$ & $t^{-5}$ & $t^{-6}$ \\
 \hline
\end{tabular}
\caption{
Scaling behavior of the $\N=8$ SUGRA and $\N=4$ SYM multi-particle unitarity cuts under the deformation defined by Eqs.~(\ref{eq:scal_deformation}) and (\ref{eq:spec_q_constraint}) for the \textbf{$\mathbf{D}$-dimensional cut integrands} up to four loops.
 }
\end{table}
An example diagram with the worst UV behavior under the specialized shift (\ref{eq:scal_deformation}) (combined with the constraint (\ref{eq:spec_q_constraint})) is 
\begin{equation}
  \raisebox{-53pt}{\includegraphics[scale=.5]{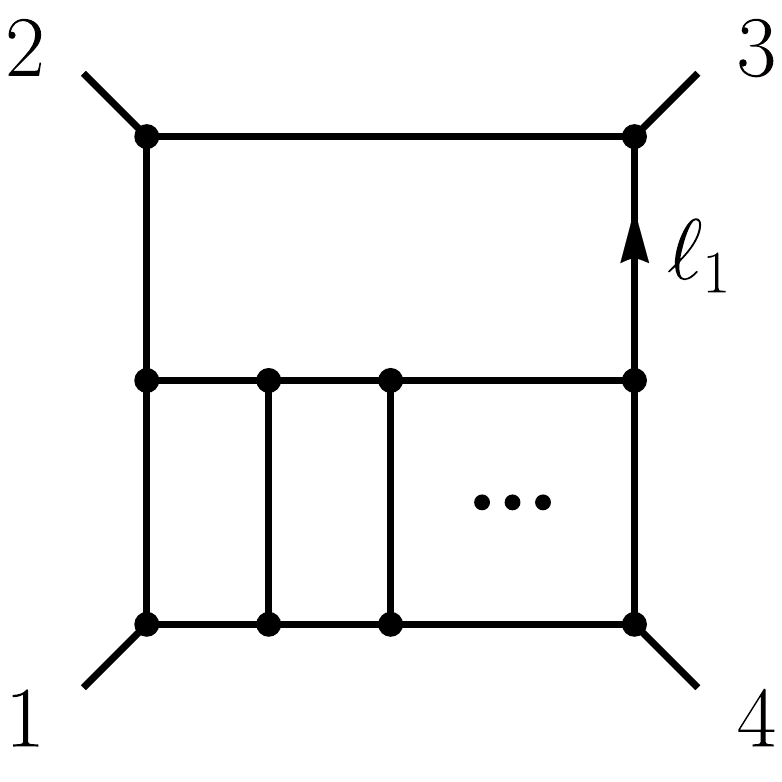}} \quad N(\ell_i,p_j) = \begin{cases}
    f_{\rm YM} \times (\ell_1\cdot p_4)^{L-2}, \quad &\N=4 \,\,{\rm SYM}\\
    f_{\rm GR} \times (\ell_1\cdot p_4)^{2(L-2)}, \quad & \N=8\,\, {\rm SUGRA}
\end{cases}
\end{equation}
On the multi-particle unitarity cut (\ref{fig:MUC}) of the diagrams in (\ref{fig:muc_diags}), there remain $3L{+}1{-}(L{+}1) = 2L$ uncut propagators, and the overall scaling of the diagram is 
\begin{equation}
\label{eq:sugra_d_scaling}
\text{SUGRA diagram scaling:} \quad \frac{(\ell\cdot p_4)^{2L-4}}{(\ell^2)^{2L}} \sim \frac{t^{2L-4}}{t^{2L}}\sim \frac{1}{t^4}\,,
\end{equation}
independent of the loop order $L$. In comparison, we find that the diagram 
behaves like $\frac{1}{t^{L+2}}$ in $\N=4$ SYM, which agrees with
the scaling of the full loop amplitude.

%=========================================================================
\subsection{Special shift in $D=4$}
\label{subsec:special_shift_4D}
%=========================================================================

Let us transition from the $D$-dimensional analysis to $D=4$, where nontrivial cancelations in cuts with
more on-shell propagators were previously identified in
\cite{Herrmann:2018dja}. In going to $D=4$, there is no change in the
scaling behavior for individual basis integrand elements.  To
reiterate, the $\N=8$ SUGRA basis elements scale like $1/t^4$, see (\ref{eq:sugra_d_scaling}), and the
$\N=4$ SYM basis elements fall off at infinity as $1/t^{L+2}$.

\begin{figure}[h]
\vskip -0.3cm
\centering
 \includegraphics[scale=1,trim={0cm 0cm 0cm 0cm},clip]{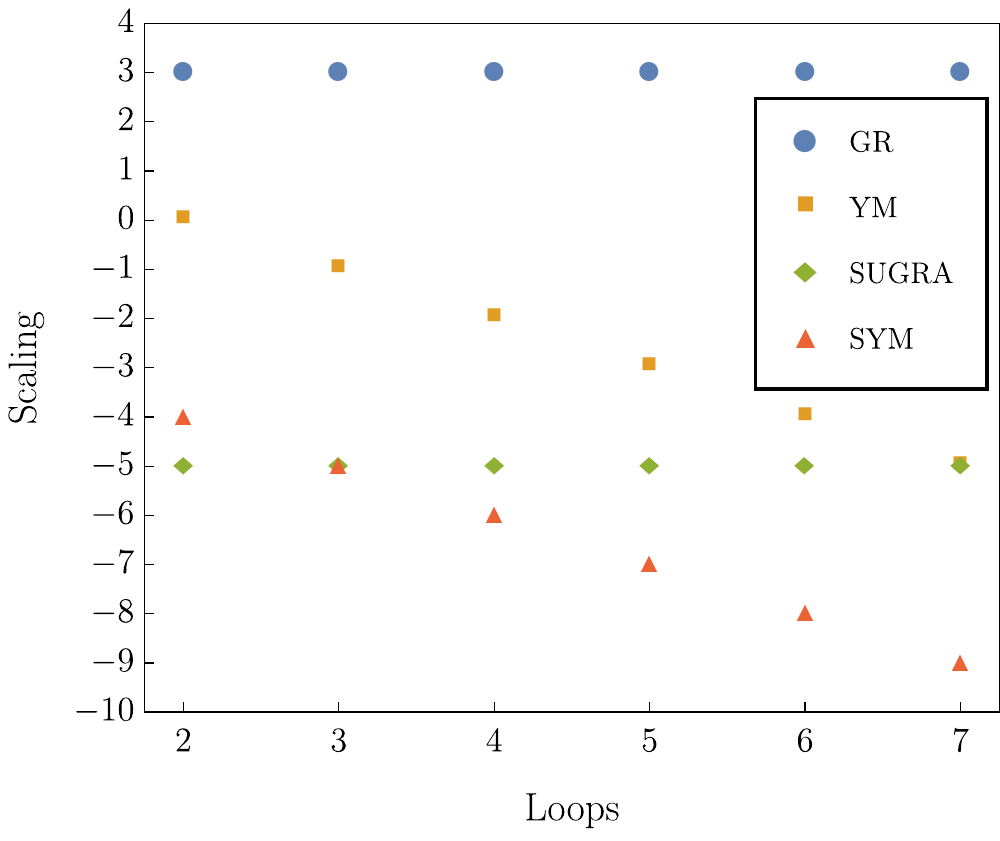}
 \vskip -.3cm
 \caption{\label{fig:uv_scaling_4d_cuts}UV scaling of $\N=8$ SUGRA, (planar) $\N=4$ SYM, 
 pure GR, and pure (planar)YM multi-particle unitarity cuts under \textbf{four-dimensional deformations} 
 with results up to seven loops. The \emph{Scaling} axis labels the leading $t$ behavior of the cuts as $t\rightarrow\infty$. The thin lines denote the scaling in D-dimensions, where the continuous part has been checked explicitly and the dashed part is conjectured.
 There is an overall improvement of one power in the large $t$ limit of gravity cuts with respect to $D$-dimensions; the same is not true for Yang-Mills. }
\end{figure}

Having analyzed individual integrals, we now perform the calculation 
for the full amplitude. Instead of starting with the integrand in terms 
of local diagrams, we use four-dimensional Yang-Mills tree-level 
amplitudes calculated via BCFW (e.g. by the package of \cite{Bourjaily:2010wh})
that are subsequently fed into the KLT relations \cite{Kawai:1985xq,Berends:1988zp,Bern:1998sv} 
to obtain gravity trees. With this setup, we compute the UV scaling results through seven loops
which are summarized in Fig.~\ref{fig:uv_scaling_4d_cuts}. We also
obtain results for the non-supersymmetric theories and get e.g. $t^3$
for GR and $1/t^{L-2}$ for YM.
%
%%%%%%%%  FIGURE %%%%%%%%
%%%%%%%%  END FIGURE %%%%%%%%

While for (super) Yang-Mills theories there is no difference, and the
$D=4$ amplitudes scale identically as their general $D$-dimensional
counterparts, in gravitational theories there is a {\bf drop by one
power}\footnote{Since Yang-Mills and gravity are closely related via KLT \cite{Kawai:1985xq,Berends:1988zp,Bern:1998sv}, it would be interesting to understand the drop in the large $t$ scaling of gravity multi-particle unitarity cuts in the $D {\to} D{=}4$ transition from this perspective.},
\begin{equation}
\label{eq:sugra_d4_scaling}
F_{\text{SUGRA}} \sim \frac{1}{t^5},
\quad 
F_{\text{GR}} \sim t^3 \qquad \mbox{for $2\leq L\leq 7$}\,.
\end{equation}
Looking more closely at the $D$ to $D=4$ transition, we find that (at
least for $L=2,3$) the leading $1/t^4$ piece of the $\N=8$ SUGRA amplitude has
the following form
\begin{equation}
F_{\text{SUGRA}} \sim \frac{\Delta}{t^4} + {\cal O}\left(\frac{1}{t^5}\right) \,,
\quad 
\mbox{where}\quad \Delta = \big(\text{Gram}[q_1q_2 \,p_1p_2p_3]\big)^2\,,
\end{equation}
and the $q_i$ in the Gram determinant denote the shift vectors of (\ref{eq:scal_deformation}). 
Importantly, $\Delta$ \emph{vanishes} in $D=4$ thereby improving the UV scaling of the amplitude to
${\cal O}(\frac{1}{t^5})$. It is worth mentioning that at the loop orders at which we performed this 
analysis the power-counting of $\N=8$ does not allow for a Gram determinant in the numerator 
of any single diagram. Crucially, many diagrams contribute to the cut and only the full sum 
assembles into the Gram determinant plus power suppressed terms at infinity. In higher loop cases, where
Gram-determinants are allowed by power-counting, further (potentially badly behaved UV terms) drop out 
in strictly four spacetime dimensions. In our four-dimensional analysis of the cuts, any such drops are taken 
into account automatically by the use spinor-helicity variables. 
Even though, we have written out the explicit form of the Gram determinant 
only for the two- and three-loop integrands, this feature is clearly behind 
the cancellation of the leading power in the UV scaling of the integrand at higher loops. 
We conclude that there is a peculiar cancelation at infinity in gravity loop integrands on multi-unitarity cuts
specifically in $D=4$ owing to the special four-dimensional kinematics.

%=========================================================================
\subsection{Comments}
\label{subsec:scaling_comments}
%=========================================================================

Studying the peculiar scaling properties of integrands at infinity begs the natural question
about the meaning of this four-dimensional feature and what
it can teach us about gravity amplitudes. We are far from having a
complete answer and currently it is difficult to relate the improved large $t$ behavior of gravity cuts 
directly to new symmetries or implications for final
amplitudes (including the status of UV divergences). However, several comments are in place.

%=========================================================================
\subsubsection*{Shift in $D=4$ and tree-level amplitudes}
%=========================================================================

Let us look at $D=4$ more closely. We choose a particular shift of
loop momenta $\ell_k \mapsto \hat{\ell}_k$ which corresponds to a
chiral shift, where the $\lamt{}$ spinors are shifted proportional to
a common reference spinor $\tw{\eta}$,
\begin{align}
\label{eq:loop_chiral_shift}
\tw{\lambda}_{\ell_k} \mapsto \widehat{\tw{\lambda}}_{\ell_k} = \lamt{\ell_k} + t\, z_k\, \tw{\eta}
\quad \mbox{for} \quad k\in\{1,\ldots,L{+}1\} \quad \mbox{subject to } \quad 
\sum^{L{+}1}_{k=1} z_k \lam{\ell_k} = 0 \,,
\end{align}
and the $\lam{\ell_k}$ remain unshifted\footnote{Note that a very similar
  shift has been discussed in the study of recursion relations for
  general $4D$ field theory tree-level amplitudes in
  Ref.~\cite{Cohen:2010mi}. In contrast to our loop-setup here,
  \cite{Cohen:2010mi} shifted all external particles with such a
  chiral shift. We thank Henriette Elvang for insightful discussions.}.  We want to understand this behavior directly in the
context of the tree-level amplitudes that enter the cut. In this case,
Eq.~(\ref{eq:loop_chiral_shift}) corresponds to a particular
multi-line chiral shift where $n-2$ legs of the tree are deformed,
\begin{equation}
\raisebox{-45pt}{
\includegraphics[scale=0.65,trim={0cm 1cm 0cm 1cm},clip]{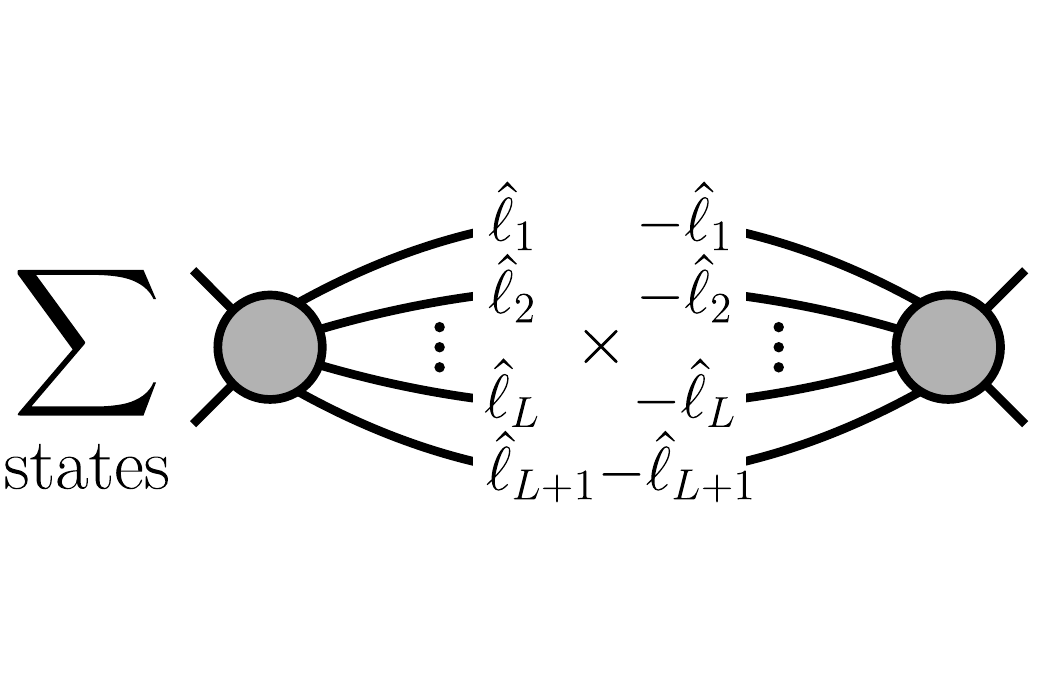}}
\end{equation}
The behavior of such deformed amplitudes for $t\rightarrow\infty$
depends on the helicities of the shifted (and unshifted) legs. The
on-shell function representing the cut of the amplitude is a product
of two tree-level amplitudes including the state sum over internal
helicities. Therefore, the individual tree-level amplitudes enter the
expression in a particular correlated way (both helicities and
shifted/unshifted momenta). For fixed internal helicities the product
of two gravity tree-level amplitudes always scales as $t^3$ or better, 
while the individual tree-level amplitudes can scale up to $t^L$
at $L$ loops, their counterpart on the other side of the cut always
compensates this poor scaling.
\begin{equation}
\raisebox{-40pt}{
 \includegraphics[scale=.55,trim={0cm .5cm 0cm 1.5cm},clip]{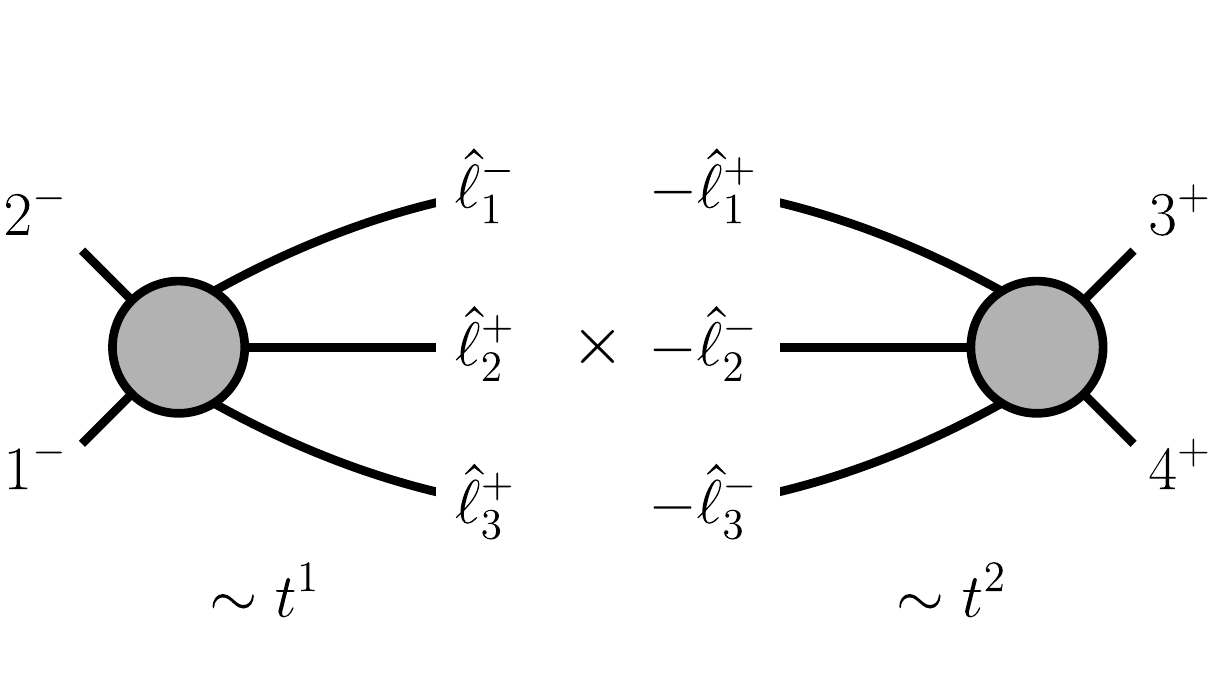}}
\end{equation}
The existence of the improved behavior of tree-level amplitudes at
infinity has been known for a very long time. The best example is the
$1/t^2$ behavior of gravity tree amplitudes under BCFW shifts, which
not only allows the reconstruction of the amplitudes from
factorizations via the BCFW recursion relations, but it also implies
the existence of bonus relations
\cite{Cachazo:2005ca,Benincasa:2007qj,ArkaniHamed:2008yf,McGady:2014lqa}. 
For generic amplitudes that fall off at infinity sufficiently fast, such
bonus relations can be recast as a sum rule on the residues of the
amplitude at finite momenta
\begin{equation}
  \A_n(t)  \sim \frac{1}{t^{2}} \quad \text{for} \quad t\to \infty \quad \longleftrightarrow \quad  
  0 = \oint\limits_{\C_{\infty}} dt\, \A_n(t) 
  = \hspace{-.3cm} 
  \underset{\begin{subarray}{c}
  	i \in\, \text{poles} \\
 	\text{of }\A_n(t)
  \end{subarray}}{\sum}  
  \hspace{-.1cm} {\rm Res}_i\,\A_n(t=t_i)
  \label{eq:bonusrelBCFW}
\end{equation}
In the supersymmetric case, our multi-line shift (\ref{eq:loop_chiral_shift}) is another example that leads to an improved
behavior of deformed amplitudes at infinity (for appropriate helicity configurations) which allows for a number of bonus relations of the type (\ref{eq:bonusrelBCFW}). More general analyses are required to
determine how the tree-level amplitudes behave at infinity for various
shifts and what are the implications for loop integrands. In
Refs.~\cite{Cohen:2010mi,Bianchi:2008pu,Cheung:2015cba,Elvang:2018dco} a number of shifts
have already been considered and we add some new data points in
Section~\ref{sec:tree_recursion}.

%=========================================================================
\subsubsection*{More cuts}
%=========================================================================

Besides the multi-particle unitarity cut described previously, there
are several other cuts with a minimal number of on-shell
propagators. In addition to permuting external legs in (\ref{fig:MUC}), 
we can also redistribute legs in the following way,
\begin{equation}
\raisebox{-22pt}{
 \includegraphics[scale=.7,trim={0cm .5cm 0cm .5cm},clip]{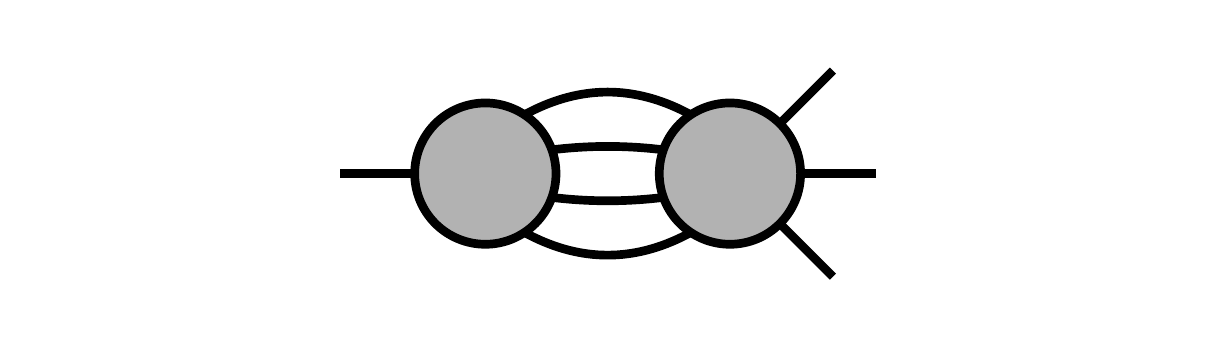}}
\end{equation}
which also has higher-point generalizations where one considers all
possible leg distributions on both sides. Apart from multi-particle
unitarity cuts we can also discuss iterated versions thereof,
\begin{equation}
\raisebox{-22pt}{
\hspace{-3.5cm}
 \includegraphics[scale=.65,trim={0cm .5cm 0cm .5cm},clip]{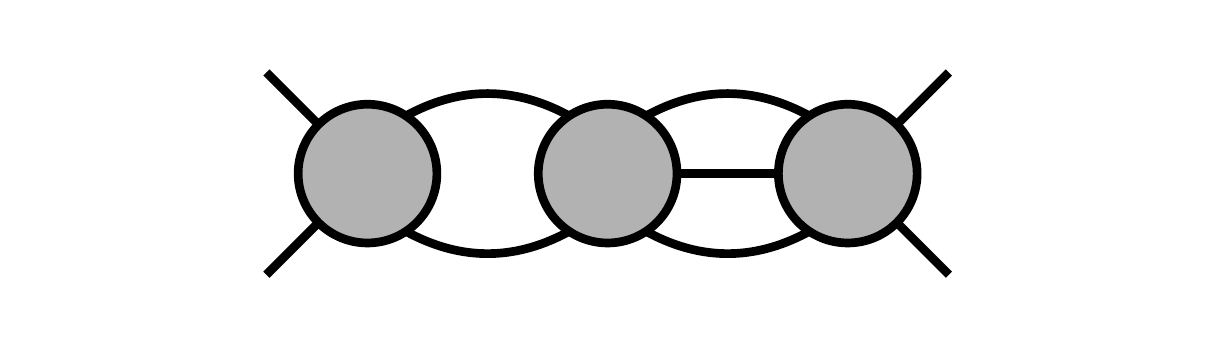}
 \hspace{-2cm}
 \includegraphics[scale=.65,trim={0cm .5cm 0cm .5cm},clip]{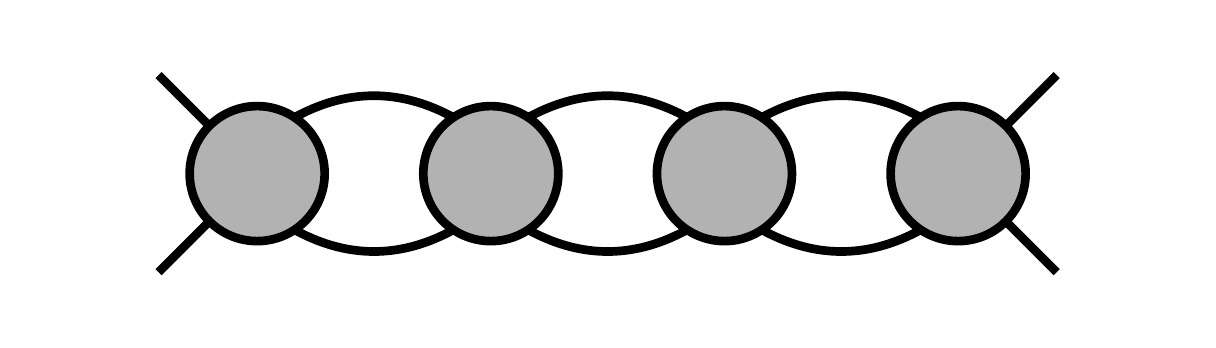}}
 \hspace{-2cm}
\end{equation}
and study their behavior at infinity under the same chiral shifts. We
indeed do find analogous drops in the large $t$ scaling specifically
in $D=4$ in all gravity theories similar to the original
multi-particle unitarity cut discussed in subsection \ref{subsec:special_shift_4D}.

Together with the earlier analysis of higher cuts
\cite{Herrmann:2018dja} it shows that these are not isolated findings
and there must exist some systematic way to capture, explain and
predict all these improved scalings in some unified way ---predicting
(rather than observing) which poles at infinity are absent, which are
present and what is the degree.

%=========================================================================

%=========================================================================
\section{Loop integrand reconstruction}
\label{sec:loop_construction}
%=========================================================================

To elaborate on the last point, we follow a particular path
explored already in the case of $\N=4$ SYM. We start with a general
ansatz for the amplitude in terms of basis integrals and impose
certain conditions trying to fix the amplitude uniquely. This ansatz procedure 
is at the heart of virtually all unitarity methods. In the most basic incarnation of 
generalized unitarity, the conditions correspond
to matching field theory on a \emph{spanning set of cuts}. In contrast, here we choose
a very special set of constraints which is inspired by a possible
geometric picture. All constraints must be {\bf homogeneous} -- meaning that 
we only impose vanishing conditions on the integrand ansatz, schematically
\begin{equation}
  \mathcal{I}_{\text{ans}} \Big|_{\text{cond.}} = 0 \, ,
  \label{eq:hom_cond}
\end{equation}
as opposed to conventional unitarity, which matches the ansatz to non-zero
functions via equations like
\begin{equation}
  \text{Cut}\left[\mathcal{I}_{\text{ans}}\right] = \sum_{\text{states}} \mathcal{A}^{\text{tree}} \times \dots \times \mathcal{A}^{\text{tree}} \, .
\end{equation}
Below, we will list more specifically, the conditions utilized in the supergravity integrand construction up to three loops. 

%=========================================================================
\subsection{Homogeneous constraints}
%=========================================================================

There are two conceptually distinct types of homogeneous constraints:

\begin{itemize}
\item Forbidden cuts:		$1a)$ field theory zeros, 
					$1b)$ helicity sector selection
\item Theory specific: constraints specific to a given theory.
\end{itemize}

Forbidden cuts refer to cuts where field theory must be zero based on general
principles, e.g. certain types of IR singularities never appear in
amplitudes, or cuts vanish for specific helicity configurations. 
An example of a constraint from category $1a)$ is a collinear cut where the loop 
momentum is proportional to an external momentum, $\ell=\alpha\, p_1$. 
Note that for gravitational theories, there are no collinear divergences \cite{Weinberg:1965nx,Akhoury:2011kq}. 
In the context of cuts, it has been shown in \cite{Herrmann:2016qea}, that gravity integrands 
vanish in all collinear regions. In more general theories, such as 
Yang-Mills, this is not the case. In those theories, from an on-shell function perspective, it is 
easy to see that loop integrands factorize
\begin{equation}
\label{eq:collinear_form}
\I^{L-\text{loop}}_n \rightarrow \frac{d\alpha}{\alpha(1-\alpha)}\times \widetilde{\I}
\end{equation}
where $\widetilde{\I}$ does not depend on $\alpha$. Therefore, the
only poles in $\alpha$ are $\alpha=0,1$ which correspond to
soft-collinear singularities making the momentum flow in propagators
$\ell^2$ or $(\ell-p_1)^2$ zero. In the on-shell diagram language, the 
$\alpha$ parameter of Eq.~(\ref{eq:collinear_form}) is associated to
the face variable of the corresponding bubble on the external leg $p_1$,
\begin{align}
\raisebox{-20pt}{
\includegraphics[scale=.6,trim={0cm .5cm 0cm .5cm},clip]{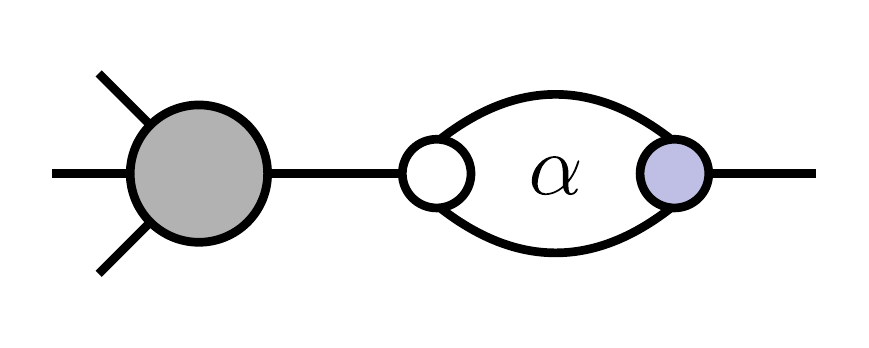}}
\end{align}
In contrast, individual integrals can have spurious collinear singularities not of the form (\ref{eq:collinear_form}) which must cancel
\begin{align}
\raisebox{-35pt}{
\includegraphics[scale=.5,trim={0cm .2cm 0cm .2cm},clip]{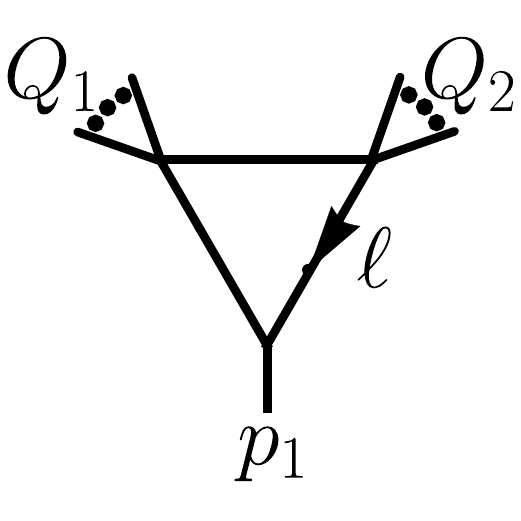}
}
\Leftrightarrow \quad  \ell^\mu = \frac{Q^2_2}{2 Q_2\cdot p_1}\, p^\mu_1 \, .
\end{align}
The cancellation can be used as an explicit constraint on an ansatz.

A simple example for a helicity-specific cut  $1b)$ appears in the context of quadruple cuts in $D=4$. 
The relevant integral topologies for MHV one-loop
amplitudes are two-mass-easy boxes. Solving the four on-shell conditions of the two-mass easy box integral 
gives two solutions. However, at the integrand level, MHV amplitudes only have nonvanishing residues on one of
them, where the three-point corners are $\bar{\text{MHV}}$ amplitudes. This cut solution enforces 
collinearity conditions on the $\lambda$ spinors of the on-shell lines. On the second solution, the MHV loop
integrand must vanish (due to $R$-charge or helicity counting) and therefore constitutes a forbidden cut.
\begin{equation}
\label{eq:forbidden_cut}
\hspace{-2cm}
\raisebox{-29pt}{\includegraphics[scale=.5]{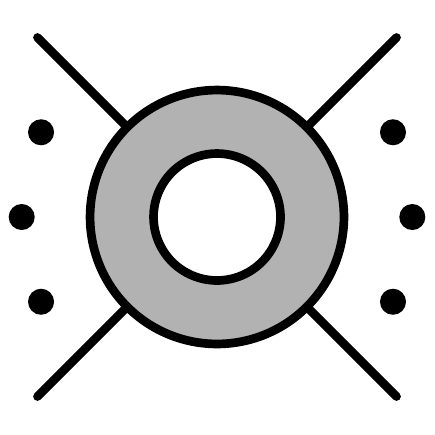}} 
\quad
\underset{\ell_j^2 = (\ell_j+p_j)^2=0}{\xrightarrow{\,\ell_i^2 = (\ell_i+p_i)^2=0}} 
\quad
\begin{cases}
  \raisebox{-30pt}{\includegraphics[scale=.5]{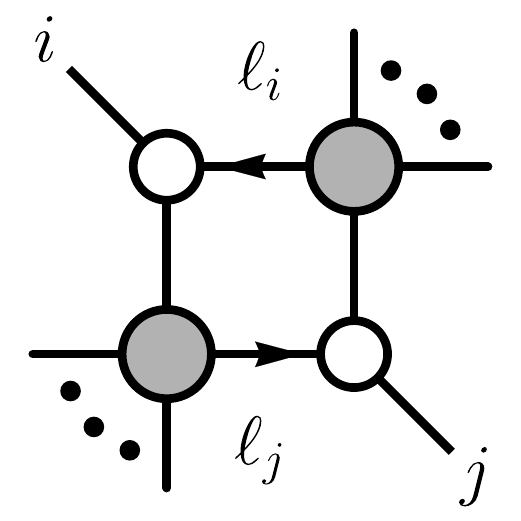}} \quad \lambda_{\ell_r} \sim \lambda_r\,, r\in\{i,j\} \quad \text{allowed,} \\
  \raisebox{-30pt}{\includegraphics[scale=.5]{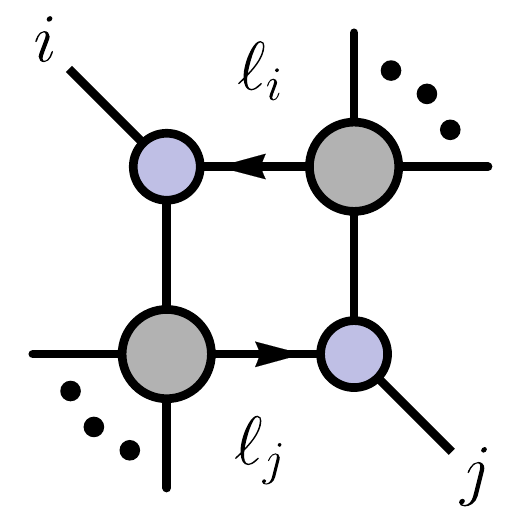}} \quad \lamt{\ell_r} \sim \lamt{r}\,,  r\in\{i,j\}\quad \text{forbidden.}
\end{cases}
\hspace{-2cm}
\end{equation}
In the context of planar $\N=4$ SYM, there are two theory-specific
constraints: the {\bf absence of poles at infinity} and
{\bf logarithmic singularities}. The first constraint corresponds to
the fact that the loop integrand never generates a singularity for
$\ell\rightarrow\infty$ anywhere in the cut structure, i.e.~there is 
never a pole (whether for real or complex $\ell$) which
localizes $\ell\rightarrow\infty$. The latter constraint is more subtle
and in momentum space it is only true for low $k$ amplitudes, 
where $k$ counts the helicity/$R$-charge of N$^{k-2}$ MHV amplitudes. 
In other cases, one can have elliptic and even more complicated singularities. 
However, if the amplitude is uplifted to bosonized momentum twistor
variables, all singularities are logarithmic and near any pole $x=0$
the loop integrand behaves as
\begin{equation}
\I\xrightarrow[]{x=0} \frac{dx}{x} \quad \mbox{where $x=f(\ell,p)$}\,.
\end{equation}
Note that this property is much stronger than just having simple poles (which is
automatic from Feynman propagators). The difference can be only
seen on higher cuts, see \cite{Bern:2014kca} for a detailed discussion. For
non-planar $\N=4$ SYM amplitudes, the same properties were conjectured to hold \cite{Arkani-Hamed:2014via},
and verified in a number of cases. However, a general proof and deeper
understanding of the theory-specific properties is still missing.

Both types of conditions alluded to above can be interpreted as the requirement that the
loop integrand vanishes on certain cuts, schematically written as
\begin{equation}
\label{eq:vanishing_cuts}
\text{Cut}_{\text{f}}\,\A_{n} = 0\,, \quad f \in \{\text{certain cuts}\}\,.
\end{equation}
For planar $\N=4$ SYM, the geometric picture for the loop integrand
directly implies that the integrand function must be fully specified
by these types of homogeneous conditions. This follows from the fact
that a positive geometry in some positive variables $x_j$ can be
defined by a set of homogeneous inequalities
\cite{Arkani-Hamed:2013jha,Arkani-Hamed:2017tmz,Arkani-Hamed:2017vfh}
\begin{equation}
\label{eq:inequalities}
h_a(x_j)\geq 0
\end{equation}
The differential form on that geometry (= loop integrand) can then be written as
\begin{equation}
\label{eq:form_def}
\Omega = \frac{N(x_j)}{\prod \,  {\rm poles}} \, \bigwedge\limits_{j=1}^{4L} dx_j\,
\vspace{-0.1cm}
\end{equation}
where the poles of $\Omega$ are dictated by the boundaries of the
geometry.  Because the numerator $N(x_j)$ is a polynomial in $x_j$, it
is fully specified by its zeroes $x^\ast_j$. Geometrically, these
zeroes correspond to special points \emph{outside} the space defined
by the inequalities in Eq.~(\ref{eq:inequalities}). Potentially, the
denominator in (\ref{eq:form_def}) can generate singularities at
locations $x^\ast_j$ where the inequalities (\ref{eq:inequalities})
are violated.  In order to not generate a spurious singularity, the
role of the numerator is to put a zero at the location $x^\ast_j$.
The crucial non-trivial statement is that in momentum space, these
$x^\ast_j$ correspond exactly to the points $f$ of the vanishing cuts
in Eq.~(\ref{eq:vanishing_cuts}) -- the denominator structure of the
loop integrand does in principle support such singularities, but the
numerators must vanish in order to prevent the appearance of a
pole. This is just a heuristic picture, which can be made more
concrete in the context of the planar $\N=4$ SYM
\cite{Arkani-Hamed:2014dca}. There, even the numerator of the form
(\ref{eq:form_def}) happens to be positive inside the positive
geometry domain suggesting a dual Amplituhedron interpretation in
which the the differential form is replaced by a volume integral.

We will not speculate further on the existence of a geometric picture
for gravity amplitudes (nonetheless, it serves as ample motivation),
but will instead investigate the ability to fully determine gravity
amplitdues imposing \emph{only} vanishing cuts
(\ref{eq:hom_cond}), (\ref{eq:vanishing_cuts}) on an ansatz.

%=========================================================================
\subsection{Amplitude reconstruction}
%=========================================================================

In this subsection, we focus on $\N=8$ SUGRA as the simplest representative of
gravitational theories, which is the most likely candidate to be fully fixed by homogeneous 
constraints. In particular, we make use of the following theory specific constraints:
\textbf{improved behavior at infinity of cuts} discussed in Sec.~\ref{sec:scaling}, and
\textbf{improved scaling of cuts under BCFW deformations} of external momenta.
We begin by constructing the two- and three-loop four-point integrands 
of $\N=8$ SUGRA.  
%=========================================================================
\subsubsection*{Two-loop four-point}
%=========================================================================
% 
We first reconstruct the integrand of the two-loop four-point amplitude from the scaling constraints at infinity.  
Originally, the integrand was calculated in \cite{Bern:1998ug} in terms of a diagrammatic expansion along the 
lines of Eq.~(\ref{eq:diag_expansion})
\begin{equation}
\label{eq:2loop4ptSUGRA}
  \raisebox{-28pt}{\includegraphics[scale=.5]{./figures/uncut2}}= \sab{12}\sab{23}\sab{13} \A_4^{\rm tree} \sum_{\sigma \in S_4} 
\left[  \frac14 \raisebox{-22pt}{\includegraphics[scale=.5]{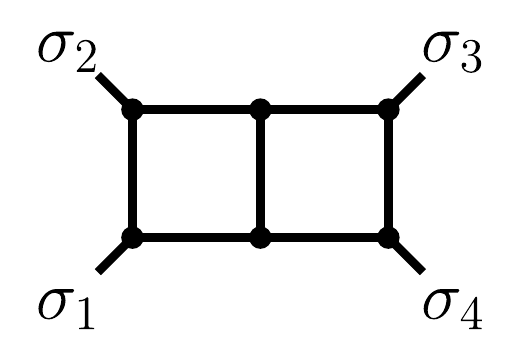}}  
+ \frac14 \raisebox{-22pt}{\includegraphics[scale=.5]{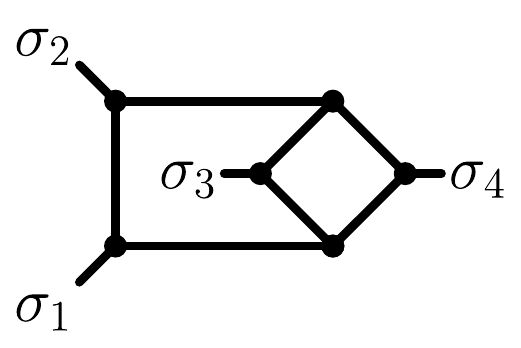}} \right]
\end{equation}
with the following numerators associated to each graph
\begin{equation}
\label{eq:2loop_nums}
   N\!\left[ \raisebox{-22pt}{\includegraphics[scale=.5]{./figures/planardb}} \right] 
= N\!\left[\raisebox{-22pt}{\includegraphics[scale=.5]{./figures/nplanardb}} \right] 
= \left(p_{\sigma_1} + p_{\sigma_2}\right)^4 =s^2_{\sigma_1\sigma_2}\,.
\end{equation}
In \cite{Herrmann:2018dja} two of the authors showed that this representation 
satisfies the improved scaling behavior at infinity when evaluated on the three-particle cut (\ref{fig:MUC}).
Instead of merely observing the consistency of the numerators (\ref{eq:2loop_nums}) with the UV scaling, 
we now demonstrate that the improved scaling conditions of Eq.~(\ref{eq:sugra_d4_scaling}) are sufficient 
to select these numerators from an ansatz.

\medskip

\begin{figure}[tb]
     \begin{center}
      \begin{tabular}{c}
	\begin{minipage}{0.3\textwidth}\begin{center}\includegraphics[width=\textwidth]{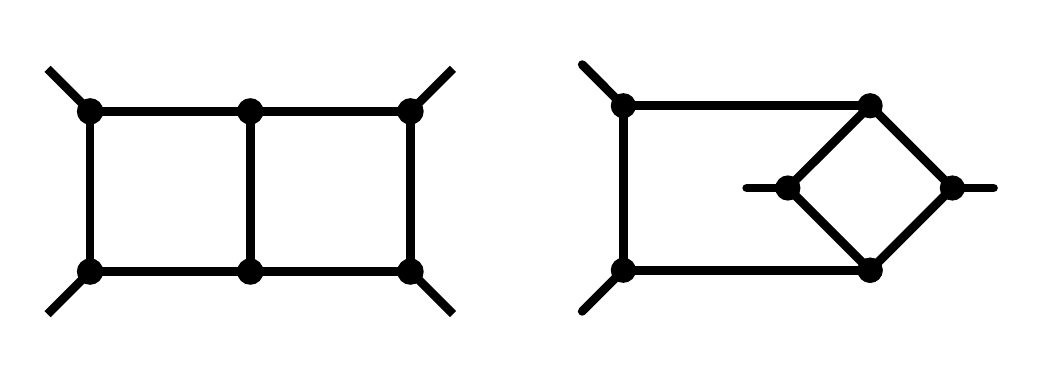}
    \end{center}
  \end{minipage}
\\
\begin{minipage}{0.75\textwidth}\begin{center}\includegraphics[width=\textwidth]{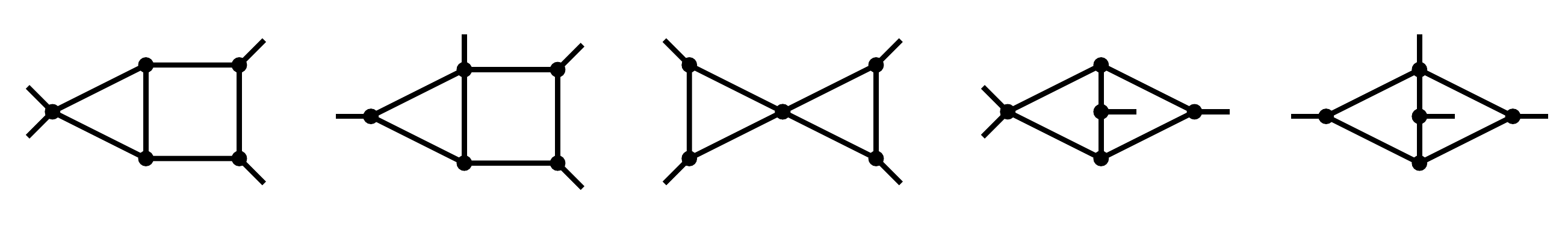}
    \end{center}
  \end{minipage}
\\
       \begin{minipage}{0.6\textwidth}\begin{center}\includegraphics[width=\textwidth]{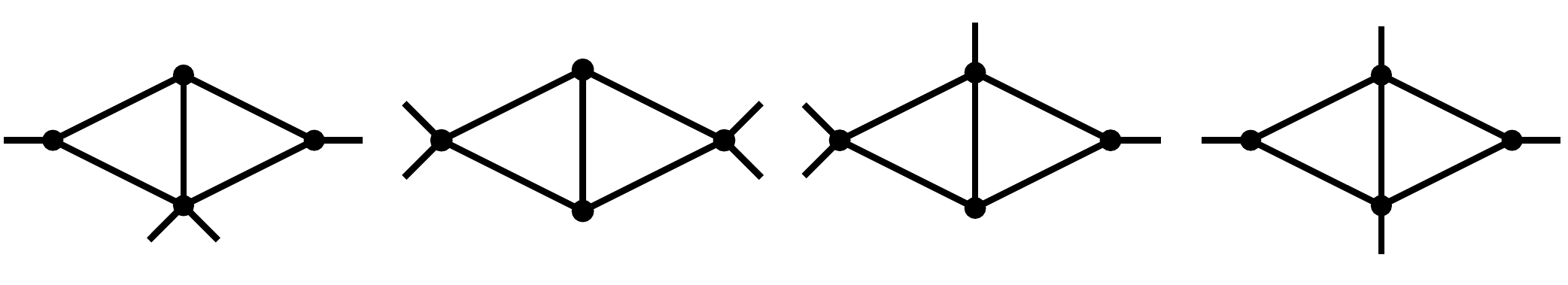}
    \end{center}
  \end{minipage}
      \end{tabular}
    \end{center}
    \caption{The integral topologies appearing in our ansatz for the two-loop four-point $\N=8$ SUGRA amplitude.}
    \label{fig:two-basis}
\end{figure}

The two-loop ansatz is built on the integral topologies shown in
Fig.~\ref{fig:two-basis}.  For each integral, we write an ansatz for its numerator with the following properties
\begin{itemize}
  \item We assign an overall factor of $\sab{12}\sab{23}\sab{13} \A_4^{\text{tree}}$ to each diagram. 
  \item We allow all terms that can be fixed purely from the maximally-supported
    cut of the diagram. All contact terms are treated as separate topologies with
    their own degrees of freedom. 
  \item We impose \emph{triangle power-counting}: we only allow numerators
    that are equivalent to scalar triangles. In particular we do \emph{not} allow
    terms of the form $(\ell_i \cdot p) (\ell_i \cdot q)$, see e.g.~\cite{Bourjaily:2017wjl} 
    for more details. Note that this is a very conservative assumption as triangle power-counting 
    is worse than what is eventually necessary for the $\N=8$ SUGRA examples discussed here.
  \item We impose diagram symmetry, that is, invariance of the numerator under all automorphisms of the skeleton graph.
  \end{itemize}

In simple cases, the numerators are composed of $\sab{ij}$ and irreducible
numerators, see e.g.~\cite{Anastasiou:2000kp}.  For more complicated diagrams,
the requirement that the numerators obey all diagram symmetries can force the
inclusion of \emph{reducible numerators} whose coefficients are however
completely locked to coefficients of irreducible ones. As such, they can be
fixed on maximal cuts. 

The two-loop planar double box for example carries a numerator ansatz
which is a degree-two polynomial built from the following scalar product building blocks
\begin{equation}
\label{eq:2l_db_eg_const}
\hspace{-1.5cm}
\raisebox{-28pt}{
\includegraphics[scale=.6]{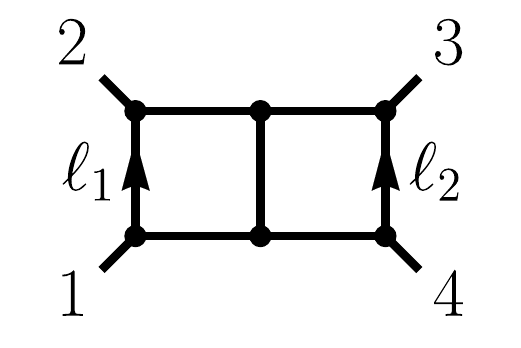}
}
\hspace{-.5cm}
\lra
\sab{12}\sab{23}\sab{13} \A_4^{\rm tree} \times
  \{\sab{12},\ \sab{23},\ p_1\!\cdot\! \ell_1,\ p_1\! \cdot\! \ell_2,\
  p_2\! \cdot\! \ell_1,\ p_2\! \cdot\! \ell_2,\ p_3\! \cdot\! \ell_1\}^2\,,
  \hspace{-1cm}
\end{equation}
where we have implicitly used momentum conservation to remove
dependence on $p_4$.  Note that this application of momentum
conservation, as well as the need for diagram symmetries, has
introduced \emph{reducible} scalar products even in this simple case.
We might expect such a numerator ansatz to have 49 free parameters.
However, imposing the symmetries and triangle power-counting reduces 
the actual degrees of freedom to 6, all of which can be in principle fixed 
on the maximal cut of the diagram.
\begin{equation}
N\!\left[
\hspace{-.35cm}
\raisebox{-28pt}{
\includegraphics[scale=.6]{./figures/planardb_1234.pdf}
}
\hspace{-.3cm}
\right]
=
\sab{12}\sab{23}\sab{13} \A_4^{\rm tree} \times
\bigg[
 c_1\, n_1 +  c_2\, n_2 + \cdots +  c_6\, n_6
\bigg]\,.
\end{equation}
The individual numerator basis elements $n_i$ can be chosen as
\begin{equation}
\label{eq:2loop_4pt_eg_num}
\begin{split}
n_1 & = \sab{12}^2\,, \quad
n_2   = \sab{12}\,\sab{23}\,, \quad 
n_3   = \sab{23}^2\,, \\
n_4 & = \sab{12} \, [ \ell_1\!\cdot (p_4- p_3)  + \ell_2\!\cdot\!(p_1 - p_2) ]\,, \\
n_5 & = \sab{23}\,[ \ell_1\!\cdot (p_4- p_3)  + \ell_2\!\cdot\!(p_1 - p_2) ]\,, \\
n_6 & = \left[ \ell_1\!\cdot (p_4- p_3)\right]\,\left[\ell_2\!\cdot\!(p_1 - p_2)\right]\,.
\end{split}
\end{equation}
Note that the basis numerators written in
Eq.~(\ref{eq:2loop_4pt_eg_num}) explicitly depend on $p_4$ for
compactness. Using momentum conservation, however, we can reduce all
dot products to the basis elements introduced in
Eq.~(\ref{eq:2l_db_eg_const}).  The remaining diagram numerators for the
rest of the potential topologies in Fig.~\ref{fig:two-basis} are
constructed in a similar manner.  Specifically, the other diagram in
the first row is also built as a degree two polynomial in the momentum
products, while the second row each carries a degree one polynomial,
and the final row diagrams are given undetermined rational
coefficients.  This ansatz contains the known integrand
(\ref{eq:2loop_nums}) by zeroing all free parameters except $c_1$ and
its counterpart in the non-planar ladder, which are set to $1$.

Next, we impose the homogeneous constraints on the ansatz
constructed as above.  The construction of four-point integrands does
not require the use of forbidden cuts to project onto the desired
helicity sector. Thus, we can solely focus on the homogeneous UV
scaling conditions. We begin by requiring the appropriate behavior at
infinity on the multi-particle unitarity cut kinematics
(\ref{eq:scal_deformation}), (\ref{eq:spec_q_constraint}).
Concretely, after calculating the cut of the ansatz, we shift the loop
momenta via (\ref{eq:loop_chiral_shift}) to get a function that
parameterizes the cut in terms of $t$
\begin{align}
  \text{Cut}_{\includegraphics[scale=.2,trim=50 0 50 0,clip]{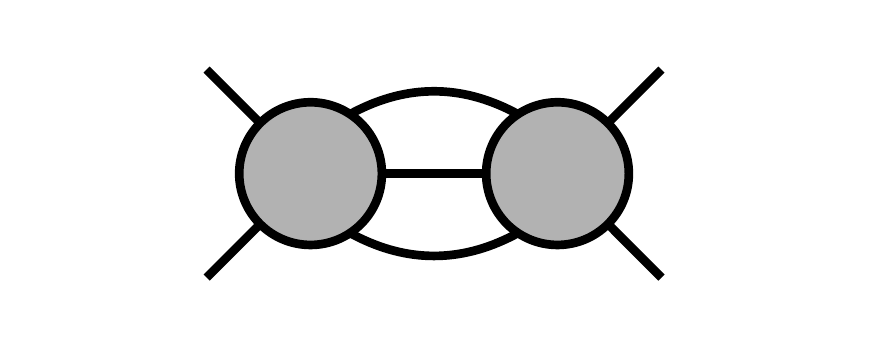}}
  \left[\mathcal{I}_{\text{ans}}\right] \xrightarrow{\text{chiral shift}} F(\{\ell,p\},t)
\end{align}
which can then be series expanded in the limit $t \to \infty$ (\ref{eq:cut_series}).  In
general this expansion of the ansatz will yield a Laurent series in $t$
\begin{equation}
  \lim_{t \to \infty} F(\{\ell,p\},t) = \sum_{i=-\infty}^{\infty} F_i(\{\ell,p\}) t^i \, .
\end{equation}
We then impose the observed scaling discussed in section
\ref{subsec:special_shift_4D}.  Specifically, we require that
\begin{equation}
  F_i(\{\ell,p\}) = 0 \quad \forall\ i > -5
\end{equation}
for generic values of $\{\ell,p\}$, from which we extract constraints
on the free parameters of the ansatz.  For the rest of this paper, we
will use shorthand of the form
\begin{equation}
 \raisebox{-28pt}{\includegraphics[scale=.6]{./figures/2loop_22}} \hspace{-1cm}
  \propto \frac{1}{t^5}\,.
\end{equation}
to denote this process of fixing parameters using cut scaling
constraints.
Enforcing this homogeneous condition determines the entire two-loop ansatz in
terms of one parameter, except for the ``kissing triangles'' topology at the center of the second row in Fig.~\ref{fig:two-basis}.  Further
consideration reveals that this is because no permutation of
such integral topology contributes to the multi-particle cut.  To resolve the missing
information, we consider an ``iterated'' two-particle cut, where we impose scaling
in one of the one-loop subdiagrams
\begin{equation}
  \raisebox{-2.3em}{\includegraphics[height=5em]{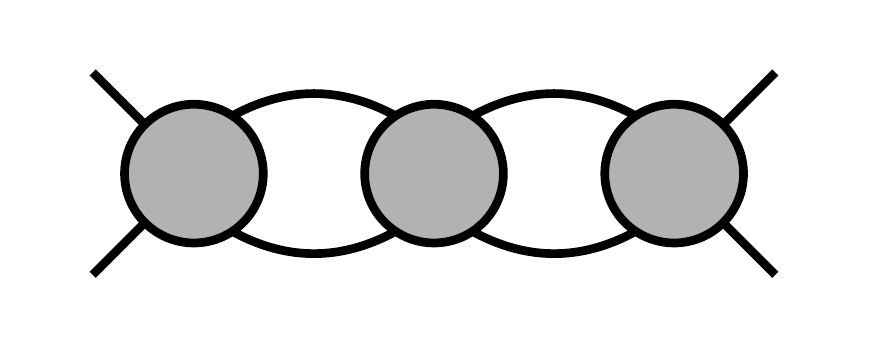}} \hspace{-.5cm} \propto \frac{1}{t^4}.
\end{equation}
Notably, the scaling is slightly different for the one-loop
subdiagram.  While we do not discuss this feature here, it is covered in detail
in the previous paper \cite{Herrmann:2018dja}. Imposing the
iterated scaling fixes the numerator for the ``kissing
triangles''.  Thus, just imposing the particular behavior at infinity,
we single out the known representation of the two-loop four-point $\N=8$ supergravity integrand in Eqns.~\eqref{eq:2loop4ptSUGRA} and \eqref{eq:2loop_nums}.

%=========================================================================
\subsubsection*{Three-loop four-point}
%=========================================================================

\begin{figure}[htb]
  \begin{subfigure}{0.5\textwidth}
    \includegraphics[scale=.6,trim={0cm .5cm 0cm .5cm},clip]{./figures/3loop_13}
    \includegraphics[scale=.6,trim={0cm .5cm 0cm .5cm},clip]{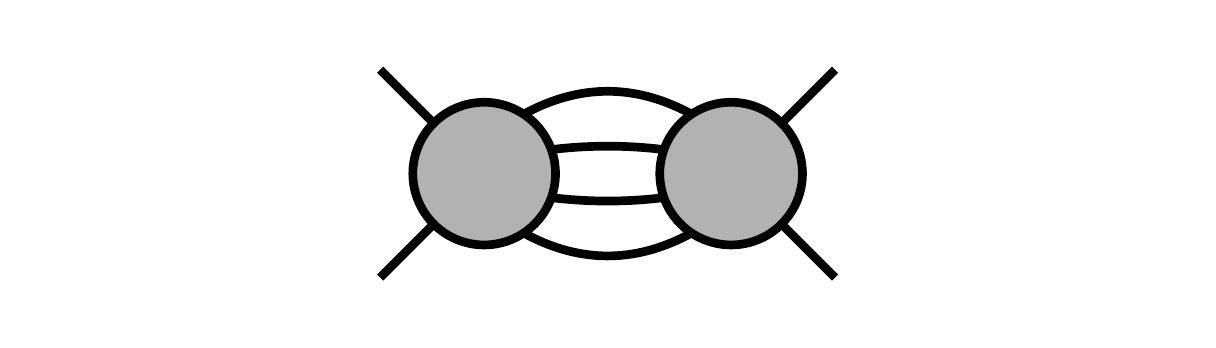}
    \caption{Three-loop mutli-unitartiy cut}
    \label{fig:three-legs}
  \end{subfigure}
  \begin{subfigure}{0.5\textwidth}
     \includegraphics[scale=.6,trim={0cm .5cm 0cm .5cm},clip]{./figures/3loop_it1}
     \includegraphics[scale=.6,trim={0cm .5cm 0cm .5cm},clip]{./figures/3loop_it2}
     \caption{Iterated three-loop cuts}
     \label{fig:three-iter}
   \end{subfigure}
   \caption{\label{fig:3L_cuts}Cut topologies considered in the UV construction of the three-loop four-point amplitude in $\N=8$ SUGRA.}
\end{figure}

At three-loop four-point, the combinatorics of the ansatz is much more
involved. After careful counting, we are left with 2758 parameters\footnote{Note that the number of parameters quoted here is even more conservative and includes a few degrees of freedom that are beyond triangle power-counting. In particular, we did not remove all terms of the form $(\ell\!\cdot\!p)(\ell\! \cdot\! q)$ from sub-boxes as we did in the two-loop analysis (see discussion below (\ref{eq:2l_db_eg_const})).} in 83
diagrams. As hinted at in the two-loop construction, to make sure all diagrams
in our basis are constrained we need to consider the scaling on a spanning set
of cuts.  Specifically, we need to consider diagrams with different
distributions of external legs as shown in Fig.~\ref{fig:three-legs}.  Additionally,
similar to the two-loop case, we need to consider the iterated cuts in
Fig.~\ref{fig:three-iter} to constrain the factorizable integrals.

Imposing the appropriate scaling for the cuts of Fig.~\ref{fig:3L_cuts}, we fix nearly all
of the terms in the ansatz.  However, there is a class of terms that
the cut scalings cannot differentiate.  For example, if we consider
the ladder diagram, we see that the ansatz for its numerator is reduced to
\begin{equation}
  N\!\left[ \hspace{-.3cm}\raisebox{-2.2em}{\includegraphics[height=5em]{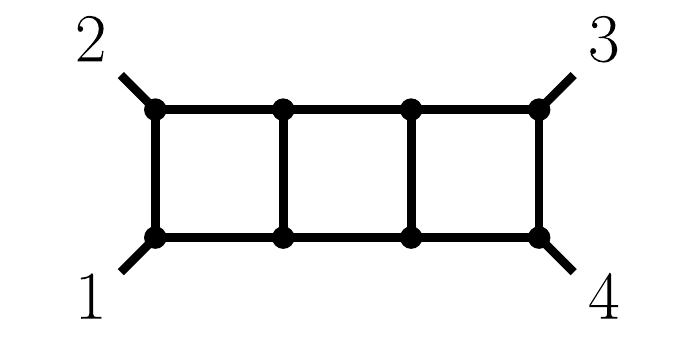}} \hspace{-.3cm}\right] 
  = a_1 \sab{12}^4 + a_2 \sab{12}^3 \sab{14} + a_2 \sab{12}^2 \sab{14}^2 \ . 
\end{equation}
Each of these terms (and similar terms in other diagrams) scales as
$t^{-6}$ on the multi-particle unitarity cut, and thus we require further
homogeneous conditions to fix them.

A natural choice for the additional constraint is to impose the appropriate
behavior of the multi-particle unitarity cut under BCFW shifts of external momenta.
In fact, we know that the on-shell function corresponding to the 
multi-particle cut behaves like
\begin{equation}
\hspace{-1.7cm}\raisebox{-2.3em}{\includegraphics[height=5em]{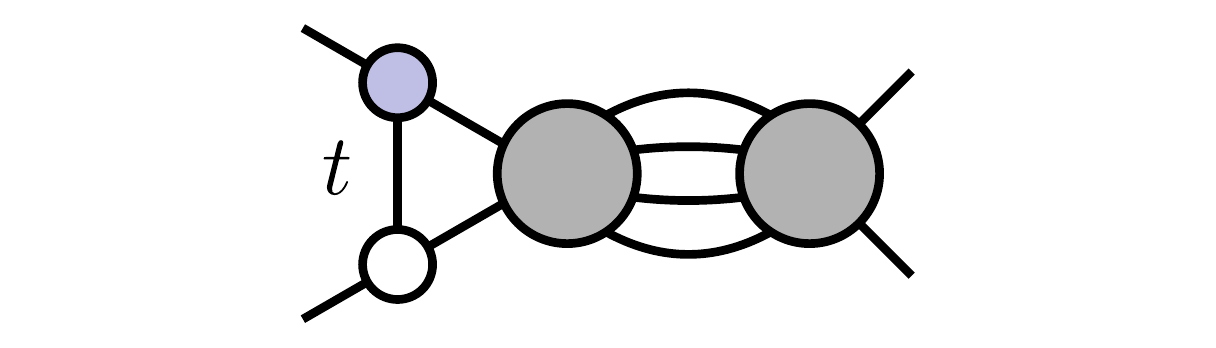}} \hspace{-1.7cm} 
\Rightarrow F(t) \sim \frac{1}{t^2} \quad\mbox{for $t\rightarrow\infty$}\,,
\end{equation}
as a consequence of the behavior of the contributing tree-level
amplitude. While all previous cuts do contain information about the
behavior at infinity of cut (on-shell) loop momenta, the last
condition also imposes constraints on the behavior at infinity for
external momenta. If we consider the union of all these constraints,
2757 of 2758 parameters get fixed, leaving us only with one overall
constant.  Thus, the scaling conditions are sufficient to fully
specify the amplitude without needing to compare with any specific
values on a cut.

The above constructions show that four-particle amplitudes in $\N=8$
SUGRA up to three loops can be fully defined by homogeneous conditions
at infinity alone.

%=========================================================================
\subsubsection*{One-loop $n$-point MHV}
%=========================================================================
A very interesting further application of the improved UV behavior is the
reconstruction of the one-loop $n$-point MHV amplitude in $\N=8$ SUGRA \cite{Bern:1998sv}. In the standard
unitarity methods the result is given by the sum of box integrals
where the coefficients correspond to leading singularities on the
quadruple cuts \cite{Britto:2004nc}. In fact, the loop integrand also includes parity odd
pentagons which integrate to zero but are needed to match all cuts
properly. A very convenient set of basis integrals are chiral boxes
\cite{Bourjaily:2013mma}.

The fact that there are no triangle and bubble integrals at one loop follows directly from
the absence of poles at infinity as was shown in \cite{Bern:2007xj}. In contrast, our construction 
starts with a complete basis ansatz for the one-loop amplitude given by box and triangle
integrals with numerators restricted by triangle power-counting. In this triangle-power-counting basis, 
the parity-odd pentagons become redundant \cite{Bourjaily:2017wjl}. Having set up the integrand basis, 
we are in the position to impose the UV scaling constraints on this ansatz. The
absence of poles at infinity on all triple cuts links the chiral box and scalar triangle numerators together. 
In fact, looking more closely at the behavior of the two particle cuts,
\begin{equation}
\label{eq:bubcut}
F(\ell) = \raisebox{-32pt}{\includegraphics[scale=.65]{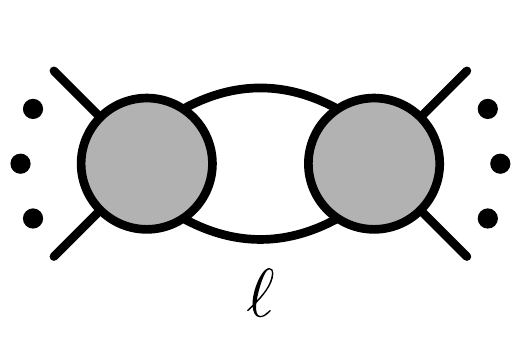}}
\end{equation}
the behavior at infinity is actually even stronger than just the absence of a pole, see the discussion in \cite{Herrmann:2018dja} for more details. In fact, this is true for any direction $\ell\rightarrow t\ell$ at infinity on the unitarity cut,
\begin{equation}
F(\ell) \sim {\cal O}\left(\frac{1}{t^3}\right)\,.
\label{eq:oneloopscaling}
\end{equation}
The mere absence of poles at infinity only requires the cut integrand to fall off like 
${\cal O}\left(\frac{1}{t^2}\right)$ as $t\to\infty$. The improved behavior in Eq.~(\ref{eq:oneloopscaling}) can in principle be used as a constraint, but does not add independent information in our one-loop MHV construction. Imposing this constraint might be necessary to construct N$^{k-2>0}$MHV amplitudes. 

As opposed to the higher loop four-point examples, imposing the vanishing of the MHV amplitude on forbidden cuts is essential. This is completely natural, for an $n$-point amplitude where one is required to specify which of the N$^{k-2}$MHV sectors one is interested in. This is easily done by demanding that our ansatz vanishes on the non-MHV cut solution of the quadruple cut (\ref{eq:forbidden_cut}). Collecting all constraints from the unitarity cuts described above for all distribution of external legs in (\ref{eq:bubcut}) we indeed fix
all coefficients in the ansatz up to an overall factor.

%=========================================================================
\subsection{(Non)-cut constructibility of ${\cal N}=8$ amplitudes}
%=========================================================================
As we saw in the previous subsection, the UV constraints are very
powerful and sufficient to completely fix the loop integrand up to
three loops at four point as well as for any number of points at one loop. While this is not an
efficient way to construct amplitudes (other methods, such as the Bern-Carrasco-Johansson 
double copy between gauge theory and gravity integrands \cite{Bern:2010ue,Bern:2019prr} 
and its generalization \cite{Bern:2017yxu}, are much more efficient), 
it shows that the $\N=8$ supergravity loop integrand can be fully
fixed (up to the orders checked and conjecturally more generally as well) using only homogeneous constraints at infinity. 
Note that we needed more than just the behavior on the
multi-particle unitarity cut (or the iterated versions of that), but also the
behavior of the cut integrand under BCFW shifts of external momenta.

This fact is not too surprising and for higher loops we  
need even more constraints at infinity. Due to gravity power-counting,
we can easily see that at a sufficiently high loop order, even at four points, there are potentially 
diagrams with the same power-counting as parent integrals, that have no propagators in one of the loops at all\footnote{In the context of dimensional regularization, such diagrams are power divergent and get set to zero. However, at the integrand level, we do not drop any such terms. For higher multiplicity, the question of non-cut-constructability has been raised previously in \cite{Bourjaily:2018omh} even for two-loop amplitudes.}. Therefore, all
multi-particle unitarity cuts (and in fact any cut that involves this loop) of this diagram vanish, and the amplitude is not cut-constructible.
\begin{equation}
\raisebox{-53pt}{\includegraphics[scale=.5]{./figures/bad_scaling_diag}} \; (\ell_1\cdot \ell_2)^{2L-6} \quad \longrightarrow \quad
\int d^4 \ell_1  \raisebox{-55pt}{\includegraphics[scale=.5]{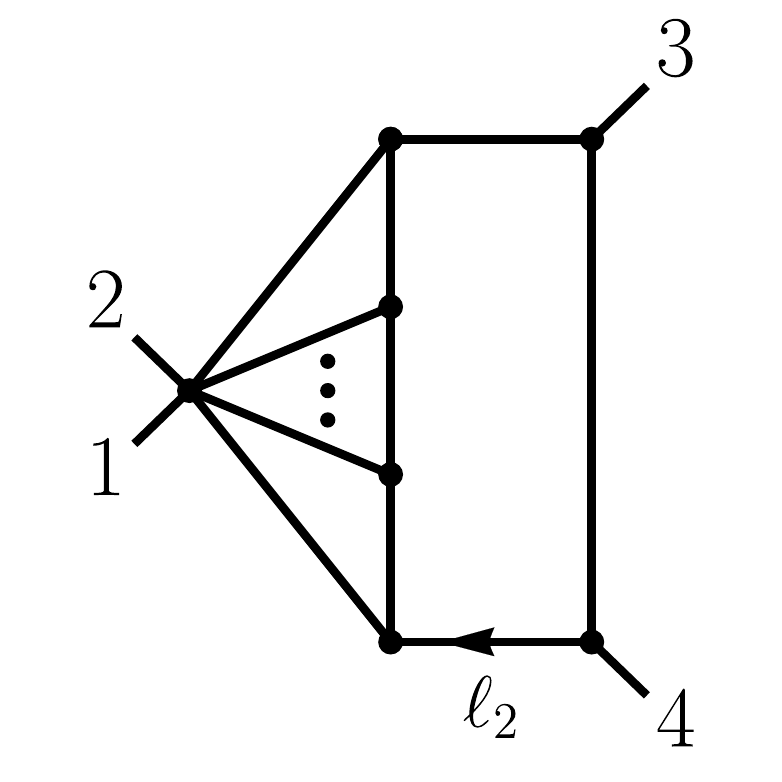}}
\label{eq:non_cut_constructable}
 \end{equation}
To illustrate this point, consider the numerator factor $(\ell_1\cdot\ell_2)$ in the left diagram of (\ref{eq:non_cut_constructable}).  
This term has the same asymptotic UV behavior as inverse propagators 
$(\ell_1{-}p_1)^2$, $\ell_1^2$, $(\ell_1{+}\ell_3)^2$, and so on. 
Therefore, starting at $L{+}1$ insertions of the numerator $(\ell_1\cdot\ell_2)$, 
we can simultaneously write down terms that completely collapse all propagator 
factors as depicted in the right diagram of (\ref{eq:non_cut_constructable}). 
This possibility first arises when $2L-6=L+1$, i.e. for $L\geq 7$. This statement 
is based on completeness properties of integrand bases \cite{Bourjaily:2017wjl} 
for a given power-counting. When considering the correct numerator of the parent 
diagram on the left of (\ref{eq:non_cut_constructable}) as dictated by the maximal 
cut, we also have to take into account the reduced diagram on the right of (\ref{eq:non_cut_constructable}) 
in our ansatz. At seven loops, the numerator ansatz includes many terms such as
\begin{equation}
N\subset\!\{(\!\ell_1\!\cdot\!\ell_2\!)^8\!,\ell_1^2(\!\ell_1{-}p_1\!)^2(\!\ell_1{-}p_{12}\!)^2(\!\ell_1{+}\ell_3\!)^2(\!\ell_1{+}\ell_4\!)^2(\!\ell_1{+}\ell_5\!)^2(\!\ell_1{+}\ell_6\!)^2(\!\ell_1{+}\ell_7\!)^2,\dots\}\,,
\end{equation}
where the second term represents the collapsed integral on the right of figure (\ref{eq:non_cut_constructable}). This diagram does not have any propagators in $\ell_1$ and therefore vanishes on all unitarity cuts. Note that kinematically there is no way to forbid such terms as they can freely mix with numerators that have non-zero contributions to cuts. Unless there is some extra constraint or mechanism which protects such integrals to appear, we have to conclude that the $\N=8$ supergravity integrands are (for sufficiently high $L$) \emph{not cut-constructible} and further conditions (apart from cuts) are required to specify the integrand uniquely. Note that after integration, terms like the one drawn on the right of figure (\ref{eq:non_cut_constructable}) vanish in dimensional regularization (power divergent terms are set to zero) and do not affect the final answer.

Similarly, as discussed in \cite{Bourjaily:2018omh}, the two-loop amplitude has poles at
infinity for $n>5$ where the degree of the pole grows with the number of
external legs. In particular, for the following integral topology,
\begin{equation}
\raisebox{-45pt}{
\includegraphics[scale=.5]{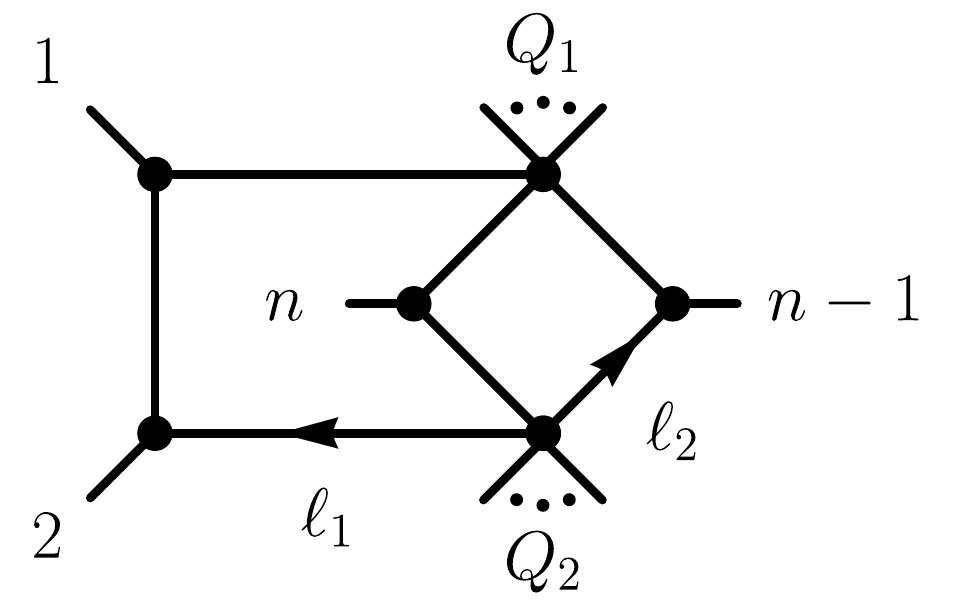}}
\end{equation}
the degree of the pole on the maximal cut forces the numerator to take the form $N = (\ell_1\cdot p)^{n-4} $. For $n\geq 12$ the ansatz would contain $(\ell_1\cdot p)^8$ as well as four inverse propagators from the $\ell_2$-loop. We are again left with pure $d^4\ell_2$ integral without any propagators. Similarly for $n\geq 14$ we can collapse the $\ell_1$-loop.\footnote{This counting is conservative and considers that $(\ell_1\cdot p)^2\sim \ell_1^2,\,\ell_2^2$. In the integral reduction we can often use relations such as $2(\ell_1\cdot p)= (\ell_1+p)^2-\ell_1^2$ and the problem might appear even for lower $n$.}
Again, the two-loop $\N=8$ amplitudes appear not cut-constructible for a sufficiently high number of points. 

We would like to point out, that these higher-loop, higher-multiplicity statements go against the ``no-triangle hypothesize'' in $\mathcal{N}=8$ SUGRA \cite{Bern:2005bb,BjerrumBohr:2006yw,Bern:2007xj,BjerrumBohr:2008ji} that has been established by explicit one-loop calculations. At the level of the analytic structure, the absence of triangles at one loop goes hand-in-hand with the absence of poles at infinity \cite{Bern:2014kca,Herrmann:2016qea,Bourjaily:2018omh}. As alluded to above, previous results of two of the authors \cite{Bourjaily:2018omh} show that even starting at two loops, at sufficiently high multiplicity, the analytic structure of the amplitudes is such that higher poles at infinity are present, which requires the introduction of triangle integrand basis elements to match these poles. 

%=========================================================================
\subsubsection*{Unification of constraints}
%=========================================================================

The presence of the no-propagator integrands requires new constraints beyond unitarity cuts.
One option is to also include constraints on the amplitude's dependence on external kinematics as a 
complement to constraints imposed by unitarity cuts. The
BCFW scaling is a natural candidate, and we already saw the successful 
application of the external kinematic shifts in fixing
the three-loop four-point amplitude.  It is obvious that only the
BCFW scaling \emph{on} multi-particle unitarity cuts can not be enough, some basis
integrals (as the one above) would directly vanish on these cuts, and
further scaling would not impose any extra constraint. Therefore, the
only possible resolution is the  \emph{simultaneous} scaling of both
external and loop momenta at infinity, basically boosting the full
amplitude to infinity. The scaling of multi-particle unitarity cuts as well as
the large $t$ scaling under BCFW shifts would then just be special cases
of this more general deformation. This also goes back to the study of the
behavior at infinity of tree-level amplitudes under general
shifts. These are very important questions and we leave them for future
work.

%=========================================================================
\section{New tree-level recursion relations}
\label{sec:tree_recursion}
%=========================================================================
%
%=========================================================================
\subsection{Helicity agnostic $(n-2)$-line shift}
%=========================================================================
%
Motivated by our discussion, we can look more closely at shifts of
graviton (pure GR, no susy) tree-level amplitudes, and explore the
behavior under various shifts in $D=4$. To reiterate the earlier
discussion of the multi-particle unitarity cut; our parametrization of
the on-shell loop momenta in Eq.~(\ref{eq:loop_chiral_shift})
corresponds to a chiral shift of $(n-2)$ legs of the tree-level
amplitude entering the cut \
\begin{equation}
\raisebox{-30pt}{
\includegraphics[scale=.5]{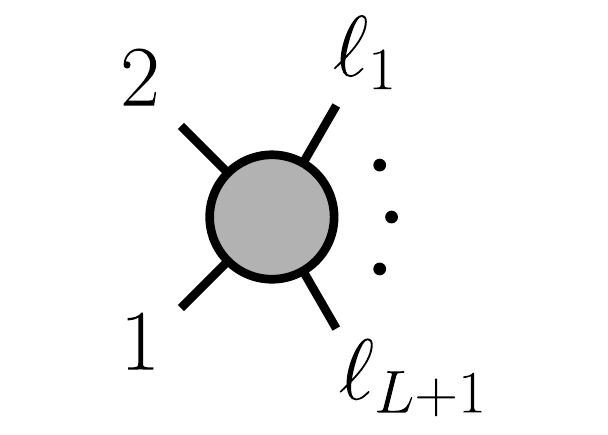}}
\longleftrightarrow
\raisebox{-30pt}{
\includegraphics[scale=.5]{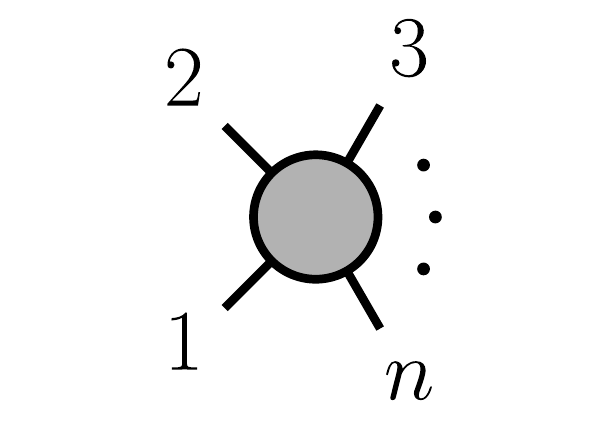}}
\end{equation}
The particular behavior for $t\to\infty$ strongly depends on the
distribution of helicities of the shifted legs. Just as in Eq.~(\ref{eq:loop_chiral_shift}), 
we shift the ``loop'' momenta in an anti-chiral fashion $\lamt{\ell_i} \to \lamt{\ell_i} + t\, z_i \tw{\eta}$ 
subject to momentum conservation which imposes a constraint on the $z_i$. 
In order to find out, whether or not an amplitude is recursively constructible by such a shift, one should study 
the asymptotic behavior of the amplitude as $t\to\infty$. Here, we try to understand the scaling properties of tree-level graviton amplitudes by studying explicit ``data''. 
\begin{table}[h!]
  \hspace{-1cm}
    \begin{tabular}{|c | c | c | c | c | }
    \cline{1-2}
      % 2-loop data
      \multicolumn{2}{|l|}{two-loop data} & \multicolumn{3}{c}{} \\
      \cline{1-2}
      $(\mi\,\mi\,\hat{\pl}\hat{\pl}\hat{\pl}) \!\sim\! t^2 $  & 
      $(\mi\,\mi\,\hat{\pl}\hat{\pl}\hat{\mi}) \!\sim\! t^1 $ & 
      \multicolumn{3}{c}{\multirow{3}{*}{}} \\
      \cline{1-2}
       $(\mi \pl \hat{\pl}\hat{\pl}\hat{\mi}) \!\sim\! t^2 $  & 
       $(\mi \pl \hat{\pl}\hat{\mi}\,\hat{\mi}) \!\sim\! t^1 $ & 
      \multicolumn{3}{c}{} \\
      \cline{1-2}
       $(\pl \pl \hat{\pl}\hat{\mi}\,\hat{\mi}) \!\sim\! t^2 $  & 
       $(\pl \pl \hat{\mi}\,\hat{\mi}\,\hat{\mi}) \!\sim\! t^{\mi7} $ & 
      \multicolumn{3}{c}{} \\
      \cline{1-3}
     \multicolumn{3}{|l|}{three-loop data}  & \multicolumn{2}{c}{} \\ 
      \cline{1-3}
      $(\mi\,\mi\hat{\pl}\hat{\pl}\hat{\pl}\hat{\pl}) \!\sim\! t^3 $  & 
      $(\mi\,\mi\hat{\pl}\hat{\pl}\hat{\pl}\hat{\mi}) \!\sim\! t^2 $ & 
      $(\mi\,\mi\hat{\pl}\hat{\pl}\hat{\mi}\,\hat{\mi}) \!\sim\! t^0 $ & \multicolumn{2}{c}{} \\
      \cline{1-3}
      $(\mi \pl \hat{\pl}\hat{\pl}\hat{\pl}\hat{\mi}) \!\sim\! t^3 $  & 
      $(\mi \pl \hat{\pl}\hat{\pl}\hat{\mi}\,\hat{\mi}) \!\sim\! t^2 $ & 
      $(\mi \pl \hat{\pl}\hat{\mi}\,\hat{\mi}\,\hat{\mi}) \!\sim\! t^0 $ & \multicolumn{2}{c}{} \\
      \cline{1-3}
      $(\pl \pl \hat{\pl}\hat{\pl}\hat{\mi}\,\hat{\mi}) \!\sim\! t^3 $  & 
      $(\pl \pl \hat{\pl}\hat{\mi}\,\hat{\mi}\,\hat{\mi}) \!\sim\! t^1 $ & 
      $(\pl \pl \hat{\mi}\,\hat{\mi}\,\hat{\mi}\,\hat{\mi}) \!\sim\! t^{\mi8} $ & \multicolumn{2}{c}{} \\
      \cline{1-4}
      \multicolumn{4}{|l|}{four-loop data}  & \multicolumn{1}{c}{} \\ 
      \cline{1-4}
      $(\mi\,\mi\hat{\pl}\hat{\pl}\hat{\pl}\hat{\pl}\hat{\pl}) \!\sim\! t^4 $  & 
      $(\mi\,\mi\hat{\pl}\hat{\pl}\hat{\pl}\hat{\pl}\hat{\mi}) \!\sim\! t^3 $ & 
      $(\mi\,\mi\hat{\pl}\hat{\pl}\hat{\pl}\hat{\mi}\,\hat{\mi}) \!\sim\! t^1 $ & 
      $(\mi\,\mi\hat{\pl}\hat{\pl}\hat{\mi}\,\hat{\mi}\,\hat{\mi}) \!\sim\! t^{\mi1} $ 
      & \multicolumn{1}{c}{} \\
      \cline{1-4}
      $(\mi \pl \hat{\pl}\hat{\pl}\hat{\pl}\hat{\pl}\hat{\mi}) \!\sim\! t^4 $  & 
      $(\mi \pl \hat{\pl}\hat{\pl}\hat{\pl}\hat{\mi}\,\hat{\mi}) \!\sim\! t^3 $ & 
      $(\mi \pl \hat{\pl}\hat{\pl}\hat{\mi}\,\hat{\mi}\,\hat{\mi}) \!\sim\! t^1 $ & 
      $(\mi \pl \hat{\pl}\hat{\mi}\,\hat{\mi}\,\hat{\mi}\,\hat{\mi}) \!\sim\! t^{\mi1} $ 
      & \multicolumn{1}{c}{} \\
      \cline{1-4}
      $(\pl \pl \hat{\pl}\hat{\pl}\hat{\pl}\hat{\mi}\,\hat{\mi}) \!\sim\! t^4 $  & 
      $(\pl \pl \hat{\pl}\hat{\pl}\hat{\mi}\,\hat{\mi}\,\hat{\mi}) \!\sim\! t^2 $ & 
      $(\pl \pl \hat{\pl}\hat{\mi}\,\hat{\mi}\,\hat{\mi}\,\hat{\mi}) \!\sim\! t^0 $ & 
      $(\pl \pl \hat{\mi}\,\hat{\mi}\,\hat{\mi}\,\hat{\mi}\,\hat{\mi}) \!\sim\! t^{\mi9} $ 
      & \multicolumn{1}{c}{} \\
      \hline
       \multicolumn{5}{|l|}{five-loop data}  \\
      \hline
      $\!\!(\mi\,\mi\hat{\pl}\hat{\pl}\hat{\pl}\hat{\pl}\hat{\pl}\hat{\pl}) \!\sim\! t^5 \!\!$  & 
      $\!\!(\mi\,\mi\hat{\pl}\hat{\pl}\hat{\pl}\hat{\pl}\hat{\pl}\hat{\mi}) \!\sim\! t^4 \!\!$ & 
      $\!\!(\mi\,\mi\hat{\pl}\hat{\pl}\hat{\pl}\hat{\pl}\hat{\mi}\,\hat{\mi}) \!\sim\! t^2 \!\!$ & 
      $\!\!(\mi\,\mi\hat{\pl}\hat{\pl}\hat{\pl}\hat{\mi}\,\hat{\mi}\,\hat{\mi}) \!\sim\! t^{0} \!\!$ & 
      $\!\!(\mi\,\mi\hat{\pl}\hat{\pl}\hat{\mi}\,\hat{\mi}\,\hat{\mi}\,\hat{\mi}) \!\sim\! t^{\mi2} \!\!$\\
      \hline
      $\!\!(\mi\pl\hat{\pl}\hat{\pl}\hat{\pl}\hat{\pl}\hat{\pl}\hat{\mi}) \!\sim\! t^5   \!\!$ & 
      $\!\!(\mi\pl\hat{\pl}\hat{\pl}\hat{\pl}\hat{\pl}\hat{\mi}\,\hat{\mi}) \!\sim\! t^4 \!\!$ & 
      $\!\!(\mi\pl\hat{\pl}\hat{\pl}\hat{\pl}\hat{\mi}\,\hat{\mi}\,\hat{\mi}) \!\sim\! t^2 \!\!$ & 
      $\!\!(\mi\pl\hat{\pl}\hat{\pl}\hat{\mi}\,\hat{\mi}\,\hat{\mi}\,\hat{\mi}) \!\sim\! t^{0} \!\!$ & 
      $\!\!(\mi\pl\hat{\pl}\hat{\mi}\,\hat{\mi}\,\hat{\mi}\,\hat{\mi}\,\hat{\mi}) \!\sim\! t^{\mi2} \!\!$\\
      \hline
      $\!\!\!\!(\pl\pl\hat{\pl}\hat{\pl}\hat{\pl}\hat{\pl}\hat{\mi}\,\hat{\mi}) \!\sim\! t^5 \!\!\!\!$  & 
      $\!\!(\pl\pl\hat{\pl}\hat{\pl}\hat{\pl}\hat{\mi}\,\hat{\mi}\,\hat{\mi}) \!\sim\! t^3 \!\!$ & 
      $\!\!(\pl\pl\hat{\pl}\hat{\pl}\hat{\mi}\,\hat{\mi}\,\hat{\mi}\,\hat{\mi}) \!\sim\! t^1 \!\!$ & 
      $\!\!(\pl\pl\hat{\pl}\hat{\mi}\,\hat{\mi}\,\hat{\mi}\,\hat{\mi}\,\hat{\mi}) \!\sim\! t^{\mi1} \!\!$ & 
      $\!\!(\pl\pl\hat{\mi}\,\hat{\mi}\,\hat{\mi}\,\hat{\mi}\,\hat{\mi}\,\hat{\mi}) \!\sim\! t^{\mi10}\!\! $\\
      \hline
    \end{tabular}
  \caption{\label{tab:scaling_summary}Scaling behavior of graviton tree-level amplitudes under the chiral deformation defined in Eq.~(\ref{eq:loop_chiral_shift}) of $(n{-}2)$ external legs. The scaling is sorted according to the different helicity configurations of the amplitudes and shifted legs are denoted by $\hat{\pm}$. The $L$-loop data corresponds to $(L+3)$-point  tree-level amplitudes.}
\end{table}
Several features stand out when looking at the large $t$ behavior in Tab.~\ref{tab:scaling_summary}: generally, reading the table horizontally (i.e. for fixed helicities of the two unshifted legs), the more negative helicity gravitons we shift by our anti-holomorphic deformation, the better the large $t$ scaling. Generally, the anti-chiral deformation of $(n-2)$ legs leads to a large $t$ behavior of the amplitude that would not allow us to recursively reconstruct the answer due to the nontrivial contribution at infinity. It is also noteworthy that the large $t$ behavior cannot be simply predicted from little group scaling and the knowledge of the mass dimension of the amplitude alone. This is in contrast to a similar all-line shift analyzed previously \cite{Bianchi:2008pu} where the behavior can be predicted. As a simple example, consider the following shifted amplitude,
\begin{equation}
\label{eq:5pt_tree_eg_shift_scaling}
\raisebox{-32pt}{
\includegraphics[scale=.5]{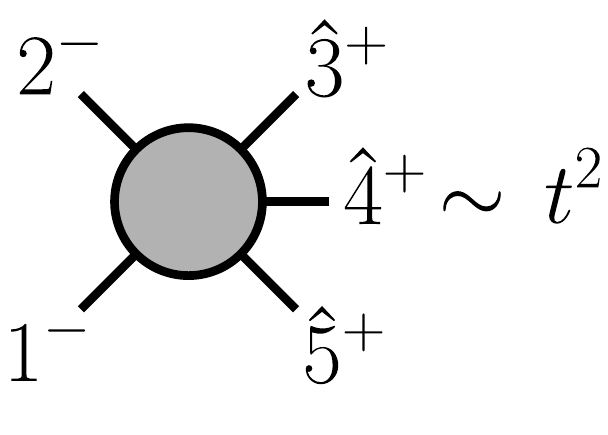}\,
}.
\end{equation}
To illustrate that the scaling of the tree amplitude does not simply follow from little group considerations, we can write down two example terms that have the same mass and little group weights, but behave very differently under the chiral shift (\ref{eq:loop_chiral_shift}),
\begin{align}
\label{eq:eg_scaling_terms}
\begin{split}
 \text{term 1:}& \quad \frac{\ab{12}^8 \ab{13}\sqb{13}\ab{23}\sqb{23}}{\ab{13}^2\ab{14}\ab{15}\ab{23}^2\ab{24}\ab{25}\ab{45}^2}  \sim t^2 \\
 \text{term 2:}& \quad\qquad\qquad \frac{\left( \sqb{34}\sqb{35}\sqb{45}\right)^2}{\sqb{12}^4}  \sim t^6
\end{split}
\end{align}
One might wonder how various representations for gravity amplitudes compare with respect to their term-wise large $t$ limit. As a first representation, let us consider the local BCJ form written in terms of cubic diagrams. For the five-particle example of Eq.~(\ref{eq:5pt_tree_eg_shift_scaling}), explicit color-Jacobi satisfying numerators are known, see appendix D of \cite{Mafra:2011kj}. Plugging these numerators into Eq.~(4.6) of \cite{Bern:2008qj} and evaluating the individual terms on the shifted kinematics (\ref{eq:loop_chiral_shift}), we see that the terms in the BCJ representation scale like $t^4$. In order to reproduce the $t^2$ behavior of the full amplitude, cancellations between different terms are therefore necessary. This is rather interesting and demonstrates once more that local representations in gravity are not ideal to faithfully represent the true UV structure of the theory.

In contrast to the BCJ representation \cite{Bern:2008qj} which is not gauge (diffeomorphism) invariant term-by-term, we can also study the KLT representation \cite{Kawai:1985xq,Berends:1988zp,Bern:1998sv} of the five-point gravity amplitude in Eq.~(\ref{eq:5pt_tree_eg_shift_scaling}). In KLT, one can expresses the full amplitude in terms of gauge invariant building blocks that, however, have spurious double poles and can schematically be written as\footnote{In all previous cases, we collectively denoted amplitudes by $\A$ representing either gravity or Yang-Mills depending on the context. Here we explicitly distinguish GR ($\M$) from YM ($\A$).} (see e.g. \cite{BjerrumBohr:2010hn,Du:2014eca})
\begin{align}
\label{eq:KLTabstract}
\M^{\text{tree}}_n = \sum_{\sigma,\rho \in S_{n{-}3}} \A^{\text{tree}}_n(1,\sigma ,n,n{-}1) S[\sigma|\rho] \A^{\text{tree}}_n(1,\rho ,n{-}1,n)\,, 
\end{align}
where $S[\sigma|\rho] $ is the momentum dependent KLT kernel and the permutation sum is over the $(n-3)!$ permutations of legs $\{2,\ldots,n-2\}$. As written in Eq.~(\ref{eq:KLTabstract}), the three legs $\{1,n{-}1,n\}$ are special, but any other choice of three legs works equally well and as we will see in a moment might be preferable at times. To be more concrete, at five points, the KLT relation reads,
\begin{align}
\label{eq:klt_5pt}
\begin{split}
\M^{\text{tree}}_5 = 	&	-\sab{12}\sab{13}\A_5(13245) \A_5(12354)
					-\sab{13}(\sab{12}{+}\sab{23})\A_5(13245) \A_5(13254) \\
				&	- \sab{12}\sab{13}\A_5(12345) \A_5(13254)
					-\sab{12} (\sab{13}{+}\sab{23})\A_5(12345) \A_5(12354)\,.
\end{split}
\end{align}
From the UV perspective, somewhat surprisingly, the KLT representation has extremely desirable properties. In fact, ``term 1'' in eq.~(\ref{eq:eg_scaling_terms}) is the worst behaved term in the KLT form of the amplitude (\ref{eq:klt_5pt}), yet scales much better at large $t$ as ``term 2'' in eq.~(\ref{eq:eg_scaling_terms}) or the BCJ pieces. In particular, one can check that the $\sab{13}\sab{23} \A_5(13245) \A_5(13254)$ term in (\ref{eq:klt_5pt}) has the same large $t$ scaling as the amplitude in Fig.~\ref{eq:5pt_tree_eg_shift_scaling} itself. This somewhat interesting observation empirically extends to all other cases we have studied. From the point of view of KLT, the behavior of the gravity amplitudes under the chiral shift is therefore inherited from the large $t$ behavior of the Yang-Mills tree amplitudes. This is even more evident when one chooses the two unshifted legs as special in the KLT formula and realizes that the KLT kernel scales uniformly at large $t$ like $t^{n-3}$, for $n\geq5$. 

For special helicity configurations where only MHV amplitudes contribute, the anti-chiral shift $\lamt{\ell_i} \to \lamt{\ell_i} + t\, z_i \tw{\eta}$ does not affect the Yang-Mills trees at all and the UV scaling comes entirely from the KLT kernel (in agreement with the scaling data in Tab.~\ref{tab:scaling_summary}). Likewise, if only $\bar{\text{MHV}}$ amplitudes are involved and the two positive helicity gravitons are taken to be $1^+$ and $2^+$, with all other gravitons having negative helicity, one can simply count the scaling of the building blocks in (a relabeled $n\lra2$ version of) Eq.~(\ref{eq:KLTabstract}): $S[\sigma|\rho] \sim t^{n-3}$, $\A^{\text{tree}}_n(1^+,\sigma ,2^+,n{-}1) \sim 1/t^n$ and $\A^{\text{tree}}_n(1^+,\rho ,n{-}1,2^+) \sim 1/t^{n-1}$ which gives the observed $1/t^{n+2}$ of this shift-sector in Tab.~\ref{tab:scaling_summary}. In the second Yang-Mills factor, one power of $t$ cancels because the two positive helicity particles are adjacent in the anti-Parke-Taylor factor and $\sqb{12}$ does not scale with $t$. For other helicity components, the analysis is much more involved but boils down to analyzing Yang-Mills tree amplitudes. 

In conclusion, it seems to be the case that the KLT representation of gravity tree-amplitudes, despite obscuring some properties (such as locality), manifests the UV scaling behavior term-by-term. It would be interesting to study this in more generality---including in $D$ dimensions.

%=========================================================================
\subsection{Same helicity $m$-line shift}
%=========================================================================
%
Even though gravity amplitudes already show an improved large $t$ behavior under the anti-holomorphic shift (\ref{eq:loop_chiral_shift}), the fall off at infinity is generically still not good enough in order to recursively construct the amplitudes. Here we study a variant of the chiral shift defined in Eq.~(\ref{eq:loop_chiral_shift})
\begin{equation}
\label{eq:chiral_shift_tree}
\vspace{-.1cm}
\lamt{j}\mapsto  \lamt{j} + t\, z_j\, \tw{\eta}
\quad \mbox{for}\quad j\subset\{1,\ldots,n\} \quad \mbox{subject to } \quad 
\sum_{j\subset\{1,\ldots,n\}} z_j \lam{j} = 0 \,,
\vspace{-.1cm}
\end{equation}
but now allow ourselves to shift \emph{any} number of legs, not just
$(n{-}2)$. The special case of a $k$-line shift (where all $k$
negative helicity particles of an N$^{k-2}$MHV amplitude are deformed)
was initially studied in \cite{Risager:2005vk} in order to derive the
CSW rules \cite{Cachazo:2004kj} in gauge theory. Applied to NMHV
amplitudes in gravity, \cite{Bianchi:2008pu} concluded that the CSW
recursion relations break down at $n=12$. In particular, the behavior
of the $n$-particle NMHV amplitude under the shift
(\ref{eq:chiral_shift_tree}) is
\begin{equation}
\label{eq:RiesagerScaling}
\M^{\text{NMHV}}_n(t) \sim \frac{1}{t^{12-n}}\,.
\vspace{-.1cm}
\end{equation}
It is obvious from the data in Tab.~\ref{tab:scaling_summary} that in order 
to get a good large $t$ behavior we can
only shift $(-)$ helicity gravitons. The dependence of the amplitude
on the $(+)$ helicity gravitons requires extra $\lamt{}$-dependent
factors in the numerator to get the correct little group
weight. Shifting these gravitons then deteriorates or spoils the large
$t$ behavior.

Consider $n$-point N$^{k-2}$MHV amplitudes with $k$ negative
helicity gravitons. We shift $m\leq k$ of these legs via the deformation 
in Eq.~(\ref{eq:chiral_shift_tree}). Based on experimental
evidence up to eight-points, we conclude that the large $t$ behavior of this $m$-line
shift is
\begin{equation}
\vspace{0cm}
\label{eq:gen_chiral_scaling}
\M^{\text{N}^{k-2}\text{MHV}}_n(t) \sim \frac{1}{t^{6+m-(n-k)}}\,.
\vspace{-.1cm}
\end{equation}
The more general scaling (\ref{eq:gen_chiral_scaling}) is in agreement
with the 3-line shift scaling of Eq.~(\ref{eq:RiesagerScaling}), as
seen by setting $m=3$, and the least favorable value $k=3$. Note that
$k=2$ is not covered by this analysis because we need at least three
negative helicity gravitons to perform the shift
(\ref{eq:chiral_shift_tree}).

The $m$-line shift provides more flexibility, and shifting a
sufficient number of external legs one can always achieve
constructibility ($\M^{\text{N}^{k-2}\text{MHV}}_n(t)$ falls off at
least like $1/t$ as $t\to\infty$). In particular, for $k>\frac{n}{2}$
we use the chiral shift (\ref{eq:chiral_shift_tree}) and the exponent
is positive for $m>(n{-}k){-}6$, i.e. if we shift more than
$(n{-}k){-}6$ negative helicity gravitons. In the extreme case when
all $(-)$ gravitons are shifted, we have $m=k$ and the amplitude
scales as
\begin{equation}
\vspace{0cm}
\M^{\text{N}^{k-2}\text{MHV}}_n(t) \sim \frac{1}{t^{6+2k-n}} < \frac{1}{t^6}\,, \qquad \mbox{for } k>\frac{n}{2}\,.
\vspace{-.1cm}
\end{equation}
If $k<\frac{n}{2}$ we can use the holomorphic chiral shift 
\begin{equation}
\lam{j} \to \lam{j} + t\, z_j \eta 
\quad \mbox{for}\quad j\subset\{1,\ldots,n\} \quad \mbox{subject to } \quad 
\sum_{j\subset\{1,\ldots,n\}} z_j \lamt{j} = 0 \,,
\vspace{-.1cm}
\end{equation}
and repeat exactly the same exercise as before. For $k=\frac{n}{2}$
both shifts are equivalent, giving a large $t$ scaling behavior of
$\M^{\text{N}^{\frac{n}{2}-2}\text{MHV}}_n(t)\sim \frac{1}{t^6}$ for
the maximal $m=k=\frac{n}{2}$ shift. In summary, we can always choose
a shift such that the behavior at infinity is at
least~$\frac{1}{t^6}$.
\vspace{-.9cm}
\newpage

%=========================================================================
\subsection*{Bonus relations}
%=========================================================================

If the deformed amplitude $\M_n(t)$ falls off at large $t$ at least
like $1/t$, we can consider a Cauchy residue theorem of the shifted
amplitude starting from a little contour around infinite $t$,
$\C_{\infty}$,
\begin{equation}
\oint\limits_{\C_{\infty}} \frac{dt}{t} \M_n(t) = 0\quad \leftrightarrow \quad \M_n(t=0) = 
- \hspace{-.3cm} 
  \underset{\begin{subarray}{c}
  	i \in\, \text{poles} \\
 	\text{of }\M_n(t)
  \end{subarray}}{\sum}  
  \hspace{-.1cm}
  \frac{1}{t_i} {\rm Res}_i\,\M_n(t=t_i)\,.
  \vspace{-.2cm}
\end{equation}

The residues at the poles $t=t_i$, $ {\rm Res}_i\,\M_n(t=t_i)$, are
then calculated as products of shifted lower point amplitudes as
usual. This is obviously not the most economic shift as the usual BCFW
shift scales like $\frac{1}{t^2}$ for any helicity configuration
(except one) and therefore is sufficient to reconstruct any tree-level
amplitude recursively. However, it is still quite interesting to see
that there is another set of shifts with even milder behavior at
infinity. This improved behavior at infinity then leads to a wider
range of bonus relations (see
\cite{Cachazo:2005ca,Benincasa:2007qj,ArkaniHamed:2008yf,McGady:2014lqa}
and the discussion in Section \ref{subsec:scaling_comments}) of the
form
\begin{equation}
\hspace{-.5cm}
  \M_n(t)  \sim \frac{1}{t^{r}} \quad \text{for} \quad t\to \infty \ \leftrightarrow \ 
  0 = \oint\limits_{\C_{\infty}} dt\, t^{r{-}2} \M_n(t) 
  = \hspace{-.3cm} 
  \underset{\begin{subarray}{c}
  	i \in\, \text{poles} \\
 	\text{of }\M_n(t)
  \end{subarray}}{\sum}  
  \hspace{-.2cm} t_i^{r{-}2} {\rm Res}_i\,\M_n(t=t_i)\,.
  \hspace{-.3cm}
  \vspace{-.3cm}
\end{equation}
It would be interesting to investigate whether imposing the behavior at infinity is enough to fix the tree amplitudes uniquely. In \cite{Rodina:2016mbk} it was shown that the $z^{-1}$ behavior of Yang-Mills tree amplitudes under BCFW shifts is enough to specify them uniquely at leading order in the soft expansion. More recently, similar UV scaling constraints were used to uniquely fix tree-level amplitudes in a variety of effective field theories \cite{Carrasco:2019qwr}. The gravity shifts explored above, in conjunction with the BCFW shift, should provide more flexibility and be part of a larger story of how graviton tree-level amplitudes behave at infinity.

The large $t$ behavior (\ref{eq:sugra_d4_scaling}) of the multi-particle unitarity cut (\ref{fig:MUC}) is inherited from the large $t$ scaling of gravity tree-level amplitudes. We saw that this behavior can be understood from KLT as a consequence of the large $t$ scaling of Yang-Mills amplitudes, and is not a property of individual BCJ terms. This suggests that Yang-Mills amplitudes are responsible for the unexpected behavior of gravity cuts. On the other hand, the improved behavior of the gravity cut was special to $D=4$, while KLT works in any spacetime dimension. Furthermore, the same cut in Yang-Mills did not improve in the $D=4$ limit, so that the improved scaling of the gravity cut (\ref{fig:MUC}) can not be explained (solely) by gauge invariance or color-kinematics duality and some new ingredient is required.

%=========================================================================
\newpage
\section{Conclusion and Outlook}
\vspace{-.3cm}
\label{sec:conclusion}
%=========================================================================
%
This work is part of a larger program to understand the
structure of scattering amplitudes at large momenta. Concretely, the goal is to understand what is
``unitarity at infinity" and how amplitudes behave in this limit. Perturbative unitarity at finite loop
momenta describes the factorization of loop integrands on poles,
closely related to discontinuities of final amplitudes on branch
cuts. No analogous statement is known about singularities located at
infinite loop momenta. In this paper, we gathered further evidence that (cuts of) gravity loop
integrands possess unexpected properties at large loop momenta. 
We have investigated this phenomenon for
multi-particle unitarity cuts through seven loops and saw that an improved 
scaling behavior is present in $D=4$ thanks to the vanishing of Gram determinants. 
This suggests that gravity amplitudes have more structure in $D=4$ than in general $D$, but further details remain to be
understood. 

The scaling properties of the integrand at infinity should also be reflected in
the structure of final amplitudes in $D=4$. As stated in the introduction,
there exists an allowed counterterm, $D^8R^4$, in $\mathcal{N}=8$ supergravity
consistent with all known symmetries of the theory (supersymmetry and duality
symmetry), relevant for a potential UV divergence at seven loops in $D=4$.  The
recent computation by Bern et al. demonstrates the presence of this counterterm
in $D_c=24/5$. However, counterterms that are allowed and present in higher
dimensions, are not necessarily present in $D=4$. A simple example is $R^4$ in
$\N=4$ sugra at 1-loop which is the relevant counterterm for the $D=8$
divergence, but it does not show up in $D=4$ at three loops
\cite{Bossard:2011tq,Bern:2012gh,Bern:2012cd}.  We do not claim that this has
to be the fate of $D^8R^4$ in $D=4$, but our on-shell analysis suggests that
$D=4$ is indeed very special from the on-shell perspective. Only in $D=4$ we
see certain integrand level UV cancellations which could hint at enhanced
cancellations after integration.

The behavior of loop integrands on unitarity cuts is directly tied to the
behavior of tree-level amplitudes, and our multi-line shift provides additional
evidence of non-trivial scaling at infinity similar to the BCFW
shift. In the last part of the paper we used the scaling properties of cuts as
homogeneous constraints to fully fix the loop integrand. Our analysis suggests 
that a more general framework to probe the behavior of amplitudes at infinity is to simultaneously shift both
(cut) loop momenta as well as external momenta, and scale them to
infinity at the same time. Our ability to fix gravity amplitudes in certain examples using only
homogeneous constraints suggests a possible geometric interpretation
similar to the planar $\N=4$ SYM case. Even if a geometric formulation 
for gravity is speculative, our analysis teaches us some important lessons about gravity, namely that the improved
behavior at infinity in fact controls the full amplitude. We also used
the same shift to construct the tree-level amplitudes using recursion
relations.

It is important to stress that the cancelations at infinity uncovered in this
paper do not seem to be a consequence of gauge invariance or supersymmetry, as
individual terms in the integrand basis with their coefficients are both
supersymmetric and gauge invariant. Furthermore, similar cancelations are
present in any two-derivative theory of gravity, supersymmetric or not. The
reason we focus on $\N=8$ supergravity is because of the availability of
explicit integrand data \cite{Bern:1998ug,Bern:2008pv,Bern:2009kd,Bern:2017ucb}
to compare against, while for pure GR the results are limited. Also, the
uniqueness construction using homogeneous data likely works only for the
maximally supersymmetric case, whereas for pure GR we have to supplement
additional information.

For planar $\N=4$ SYM in $D=4$, the (complete)
absence of poles at infinity is a consequence of dual conformal
symmetry \cite{Drummond:2007au,Drummond:2006rz}. The same property was
conjectured to be true for the full (non-planar) $\N=4$ SYM theory
suggesting there is a hidden symmetry in the full theory too. First steps in this
direction have been pursued in \cite{Bern:2014kca,Bern:2015ple}, see
also interesting related work \cite{Bern:2017gdk,Bern:2018oao,Chicherin:2018wes}. 

We have seen earlier that gravity loop integrands do have poles at
infinity, as demonstrated on maximal cuts. Therefore, no analogue of
dual conformal symmetry can be present. On the other hand, the poles
at infinity surprisingly cancel in certain directions when approaching
infinity in $D=4$. Our observations fall
into the same category as the large $z$ behavior of amplitudes under
BCFW shifts, and provide further evidence that something is missing in
our understanding of gravity amplitudes. Following the traditional
logic that the properties of the S-matrix are a consequence of
symmetries, it is suggestive that this phenomenon is indeed caused by
some yet-to-be found symmetry or some novel property of general
relativity. %On the other hand, we can not exclude that more conservative explanation is possible.

%=========================================================================
\vspace{-.5cm}
\section*{Acknowledgements}
\vspace{-.3cm}
%=========================================================================
We are grateful to Zvi Bern, Henriette Elvang, and Chia-Hsien Shen for enlightening discussions and comments on the manuscript.
A.E.\ and J.P.-M.\ thank the Mani
L. Bhaumik Institute for generous support.
J.P.-M.\ also thanks SLAC for hospitality.
J.P.-M.\ is supported by the U.S.  Department of
State through a Fulbright Scholarship. 
E.H.\ is grateful to the Mani L. Bhaumik Institute for Theoretical Physics at UCLA, for hospitality during various stages of this project, 
and the Aspen Center for Physics, which is supported by National Science Foundation grant PHY-1607611.
The research of A.E. is in part supported by the Knut and Alice Wallenberg Foundation under KAW 2018.0116, {\it From Scattering Amplitudes to Gravitational Waves.}
The work of E.H.\ is supported by the U.S. Department of Energy (DOE) under contract DE-AC02-76SF00515.
The research of J.T. is supported in part by U.S. Department of Energy grant DE-SC0009999 and by the funds of University of California.

%=========================================================================
%Bibliography - from "References.bib"
%=========================================================================

\bibliographystyle{JHEP}
\phantomsection         %to get hyperlinks right for index
            %\addcontentsline{toc}{section}{\numberline{}References}
						% If you want to put a References section into the toc just comment in 
						% the previous line!
            \bibliography{gravity_at_UV.bib}
            %\printbibliography[prenote=preNote,postnote=postNote]
            \clearpage

\end{document}